\documentclass[twocolumn]{aastex63}

\usepackage{amsmath}

\newcommand{\msun}{{M_\odot}}

\newcommand{\vesc}{v_{\rm{esc}}}
\newcommand{\rperi}{r_{\rm{peri}}}

\newcommand{\vmin}{v_{\infty,\rm{min}}}
\newcommand{\bin}{\rm{bin}}
\newcommand{\tot}{{\rm{tot}}}

\newcommand{\wide}{\rm{w}}

\newcommand{\cosmic}{{\tt{COSMIC}}}
\newcommand{\gaia}{{\em{Gaia}}}
\newcommand{\vinf}{v_{\infty}}
\newcommand{\au}{{\rm{au}}}
\newcommand{\days}{{\rm{days}}}
\newcommand{\gyrs}{{\rm{Gyr}}}
\newcommand{\hrs}{{\rm{hr}}}

\newcommand{\melmwd}{{M_{\rm{EW}}}}
\newcommand{\mpert}{{M_{\rm{s}}}}
\newcommand{\porb}{{P_{\rm{orb}}}}
\newcommand{\vorb}{{v_{\rm{orb}}}}
\newcommand{\kms}{{\rm{km\,s^{-1}}}}
\newcommand{\fewbody}{{\tt{Fewbody}}}
\newcommand{\bse}{{\tt{BSE}}}
\newcommand{\ahs}{{a_{\rm{hs}}}}
\newcommand{\bmax}{{b_{\rm{max}}}}
\newcommand{\blast}{{b_{\rm{last}}}}
\newcommand{\vrecoil}{{v_{\rm{recoil}}}}
\newcommand{\ini}{{\rm{ini}}}
\newcommand{\fin}{{\rm{fin}}}
\newcommand{\mcl}{{M_{\rm{cl}}}}
\newcommand{\rcl}{{R_{\rm{cl}}}}
\newcommand{\ncl}{{N_{\rm{cl}}}}
\newcommand{\fcl}{{f_{\rm{cl}}}}
\newcommand{\sigmaall}{{\sigma_{\rm{all}}}}
\newcommand{\sigmawide}{{\sigma_{\rm{w}}}}
\newcommand{\sigmaejected}{{\sigma_{\rm{ej}}}}
\newcommand{\sigmawideejected}{{\sigma_{\rm{w},\rm{ej}}}}

\newcommand{\Rwide}{{R_{\rm{w}}}}

\newcommand{\Rwideejected}{{R_{\rm{w},\rm{ej}}}}
\newcommand{\Gammaall}{{\Gamma_{\rm{all}}}}
\newcommand{\Gammawide}{{\Gamma_{\rm{w}}}}

\newcommand{\Gammawideejected}{{\Gamma_{\rm{w},\rm{ej}}}}
\newcommand{\plummerscale}{{\mathfrak{b}_0}}
\newcommand{\eobs}{{e_{\rm{obs}}}}

\shorttitle{Forming Wide-Orbit EW Binaries}

\newcommand{\ab}{{KIC 8145411 system}}
\begin{document}

\title{Dynamically Forming Extremely Low-Mass White Dwarf Binaries in Wide Orbits}
\email{ambreesh.khurana@tifr.res.in}
\author[0000-0003-3259-6702]{Ambreesh Khurana}
\affil{Department of Astronomy \& Astrophysics, Tata Institute of Fundamental Research, Homi Bhabha Road, Navy Nagar, Colaba, Mumbai 400005, India}
\author[0000-0001-9685-3777]{Chirag Chawla}
\affil{Department of Astronomy \& Astrophysics, Tata Institute of Fundamental Research, Homi Bhabha Road, Navy Nagar, Colaba, Mumbai 400005, India}
\email{souravchatterjee.tifr@gmail.com}
\author[0000-0002-3680-2684]{Sourav Chatterjee}
\affil{Department of Astronomy \& Astrophysics, Tata Institute of Fundamental Research, Homi Bhabha Road, Navy Nagar, Colaba, Mumbai 400005, India}

\begin{abstract}

The detection of a $0.2\,\msun$ extremely low-mass white dwarf (EW) in a wide orbit ($\porb\approx450\,\days$) with a $1.1\,\msun$ main-sequence companion KIC 8145411, challenges our current understanding of how EWs form. The traditional channel for EW formation, via mass transfer from the EW's progenitor, is expected to form EW binaries in tight orbits. Indeed, the majority of known EWs are found in tight binaries with a median $\porb\approx 5.4\,\hrs$. Using numerical scattering experiments, we find that binary-binary strong encounters in star clusters can sufficiently widen the orbit of a typical EW binary, to explain the observed wide orbit of the \ab. The $\porb$ distribution for EW binaries produced through binary-binary encounters is bimodal: one mode corresponds to the initial orbital period of the EW binary, while the other is near $\porb\sim$ few $10^2\,\days$, similar to the orbital period of the \ab. We find that the production of wide EW binaries that are also ejected from the cluster peaks at a star cluster mass of $\sim10^5\,\msun$ with a rate of $\sim10^{-3}\,\rm{Gyr^{-1}}$. Assuming that $50 \%$ of all stars form in star clusters and an initial cluster mass function $\propto m^{-2}$, we estimate a galactic formation rate of $\sim4.16\times10^3\,\rm{Gyr^{-1}}$ for wide EW binaries.

\end{abstract}

\section{Introduction}
\label{sec:intro}

The discovery of an extremely low-mass ($0.2\,\msun$) white dwarf (EW) in an unusually wide orbit ($\approx 1.27\,\au$) around a main-sequence (MS) star, KIC 8145411 \citep{Masuda2019}, challenges traditional formation theories for EWs. While many EWs have already been observed \citep[e.g.,][]{Kosakowski2020,MataSanchez2020,Brown2022} with masses similar to or even lower than that of the EW in the \ab, its orbital period ($\porb\approx450\,\days$) is more than three orders of magnitude longer than the typical value of that for observed EW binaries \citep[median $\porb\approx 5.4\,\hrs$;][]{Brown2016}, making it a very interesting object.

While it may be possible for some metal-rich ($\rm{[Fe/H]}\gtrsim0.4$) isolated single stars to form low-mass ($\sim0.45\,\msun$) white dwarfs (WDs) in isolation, due to severe mass loss through stellar winds, which may limit core growth below the He-burning limit \citep{Kilic2007}, the EWs with mass $\lesssim0.2\,\msun$ are expected to form only in tight binaries. This is because the universe is not old enough for an isolated star that would leave such a low-mass remnant to evolve to a WD.
The most accepted channel for EW formation is that a close companion strips off the outer envelope of the EW progenitor while the latter is on the red-giant branch (RGB), dramatically limiting core growth. As a result, the unusually low-mass core cannot ignite helium as it approaches the asymptotic giant branch \citep[e.g.,][]{Marsh1995,Sun2018,Li2019}. While the details of the mass-transfer process can be complex, depending on which binaries go through stable mass transfer (SMT) versus common-envelope (CE) evolution \citep[e.g.,][]{Sun2018,Li2019}, it is generally accepted that the formation of an EW involves mass loss from the envelope of the progenitor star via Roche-lobe overflow (RLOF) in a tight binary. Most recently, using the location on \gaia's color-magnitude diagram and short-period $(<6\,\hrs)$ ellipsoidal variability in the ZTF light curves, \citet{El-Badry2021a} identified binaries that are on the verge of ceasing or have recently ceased mass transfer. These systems are thought to bridge 
the gap between cataclysmic variables and EWs, and the donors in these systems are thought to be potential progenitors of EWs. 

 Theoretical studies have shown that if the origin of EWs is indeed SMT via RLOF, then the mass of the EW ($\melmwd$) is expected to be related to the $\porb$ of the EW binary \citep[e.g.,][]{Rappaport1995,Tauris1999,Lin2011,2014A&A...571A..45I,Istrate2016}. 
 According to this channel, the progenitor of the EW, initially the primary of the binary, first overflows its Roche lobe during its ascent to the RGB or near the end of the MS. This leads to a phase of SMT severe enough to turn the donor into the secondary. The mass transfer stops when the donor, now the secondary, becomes smaller than its Roche lobe, due to expansion of the orbit, thus freezing the $\porb$ \citep{1971A&A....13..367R}. As a result, the final $\porb$ is a function of the radius of the giant at the end of the mass-transfer phase. On the other hand, the radius of a giant with a well-developed core depends largely on its core mass \citep[e.g.,][]{1971A&A....13..367R,Rappaport1995}. Consequently, the final $\porb$ becomes dependent on the core mass. This dependence translates to a relation between $\porb$ and $\melmwd$ since the progenitor's core eventually evolves to become the EW. 
Instead of SMT, if RLOF leads to CE evolution, then the resulting $\porb$ is much smaller compared to the limit expected from SMT \citep[e.g.,][]{Li2019}. 

\begin{figure}
    \epsscale{1.2}
    \plotone{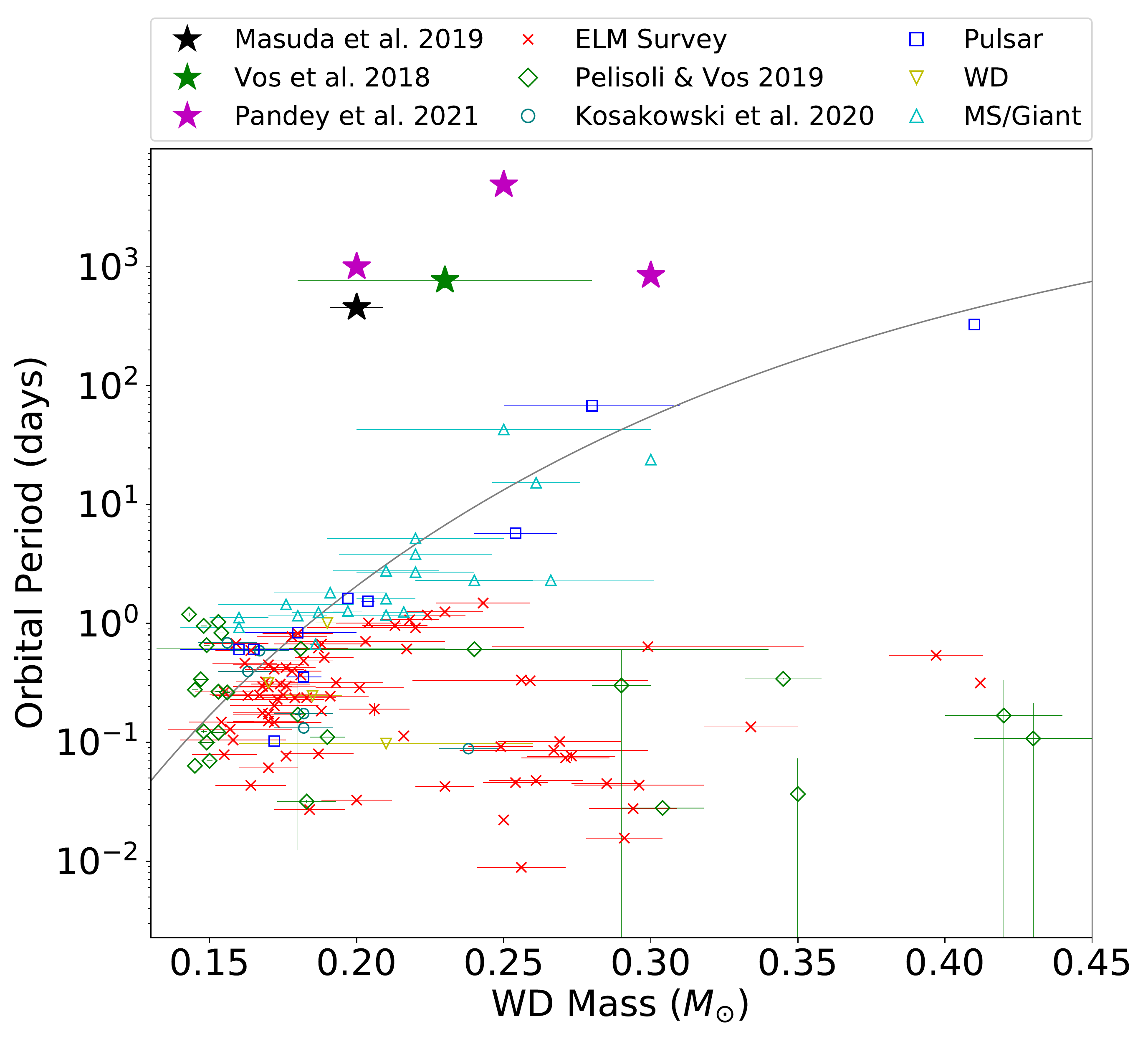}
    \caption{WD mass vs. orbital periods for EW binaries and potential progenitors compiled from various observational works. The gray curve shows the SMT boundary of \citet{Lin2011}. Any binary significantly above this curve cannot be explained by the traditional mass-transfer channel for EW production. The black star marks the \ab{} \citep{Masuda2019}. The green and magenta stars mark the other interesting candidates in wide orbits \citep{Vos2018,Pandey2021}. The red crosses and teal circles denote the ELM survey \citep{Brown2020,Brown2022} and the ELM survey south \citep{Kosakowski2020}, respectively. The candidates compiled by \citet{Pelisoli2019} are shown by the green diamonds. Finally, the blue squares, yellow inverted triangles, and cyan upright triangles represent other EW candidate binaries with pulsar, WD, and MS/giant companions, respectively.
    }
    \label{fig:survey}
\end{figure}

In \autoref{fig:survey}, we present an updated version of the Figure\ 5 from \citet{Masuda2019}, showing the WD mass and $\porb$ for observed EW binaries and possible progenitors (pre-EW binaries) collected primarily from the latest Extremely Low-Mass (ELM) survey \citep{Brown2020,Brown2022}, the ELM survey south \citep{Kosakowski2020}, and a table compiled by \citet{Pelisoli2019}, in addition to several other individual detections \citep[][]{VanKerkwijk1996,2002ApJ...570L..89D,Jacoby2005,Splaver2005,Bassa2006,Kilic2007a,Kilic2007b,Verbiest2008,Vennes2009,Vennes2011,VanKerkwijk2010,Antoniadis2012,Breton2012,Corongiu2012,Pietrzynski2012,Antoniadis2013,Maxted2013,Ransom2014,Tauris2014,Faigler2015,Rappaport2015,Guo2017,Lee_2017,Zhang2017,Brogaard2018,Lee2018,Vos2018,Jadhav2019,Ratzloff2019,MataSanchez2020,Wang2020,Pandey2021} using spectroscopic studies, pulsar timing, and photometric variability. The companions of EWs in the ELM survey are expected to be primarily CO WDs with perhaps some neutron stars \citep[][]{Andrews2014,Boffin2015}. The gray curve indicates the boundary expected from SMT \citep[][]{Lin2011}. Clearly, most observed EW binaries have $\porb$ near the boundary or well below it.

In contrast, the \ab, denoted by the black star, has a $\porb$ that places it significantly above the boundary expected from the mass-transfer channel of formation. The relevant observed properties of the \ab\ are summarized in \autoref{tab:ab}. In addition to having an orders-of-magnitude larger $\porb$ than expected, the \ab\ also has an MS star as a companion. A few other interesting systems with large $\porb$ have been discovered; these include a pre-EW in a binary with $\porb\approx771\,\rm{days}$ \citep[denoted by the green star;][]{Vos2018} and three candidate EWs with blue straggler companions identified in the star cluster M67 \citep[denoted by the magenta stars;][]{Pandey2021}. 

In this paper, we investigate the possible formation channels of EW binaries in wide orbits with similar properties to the \ab,\ without invoking any novel binary stellar evolution mechanism. First, we investigate whether normal binary stellar evolution can produce such a system (\autoref{sec:popsynth}). We then investigate whether binary-mediated strong encounters, expected to happen frequently in the cores of dense star clusters, can dynamically alter typical short-period EW binaries  \citep{Brown2016} to produce the observed \ab.  
In \autoref{sec:theory}, we investigate the energetics of the problem and show that, at the simplest level, binary-binary interactions are required for the production of EW binaries in wide orbits similar to the \ab. 
We describe the setup of our numerical simulations in \autoref{sec:method} and present the key results in \autoref{sec:results}. We summarize and discuss in \autoref{sec:conclusions}.

\begin{deluxetable}{lc}
\tablecolumns{2}
\tabletypesize{\footnotesize}
\tablecaption{Observed properties of the \ab\ \citep[Last Column of Table 2,][]{Masuda2019}\label{tab:ab}}
\tablehead{
 \colhead{Property} & \colhead{Value}}
\startdata
 Mass of the MS star, $M_{\rm{MS}}\ (\msun)$ & $1.132_{-0.078}^{+0.078}$ \\
 Mass of the EW, $M_{\rm{EW}}\ (\msun)$ & $0.200_{-0.009}^{+0.009}$ \\
 Orbital period, $P$ (days) & $455.826_{-0.011}^{+0.009}$ \\
 Semi-major axis, $\rm{SMA}$ (au) & $1.276_{-0.028}^{+0.027}$ \\
 Eccentricity, $e$ & $0.143_{-0.012}^{+0.015}$ \\
 Metallicity of the MS star, [Fe/H] & $0.39_{-0.09}^{+0.09}$
\enddata
\end{deluxetable}

\section{Binary stellar evolution}
\label{sec:popsynth}

\begin{figure}
    \centering
    \epsscale{1.2}
    \plotone{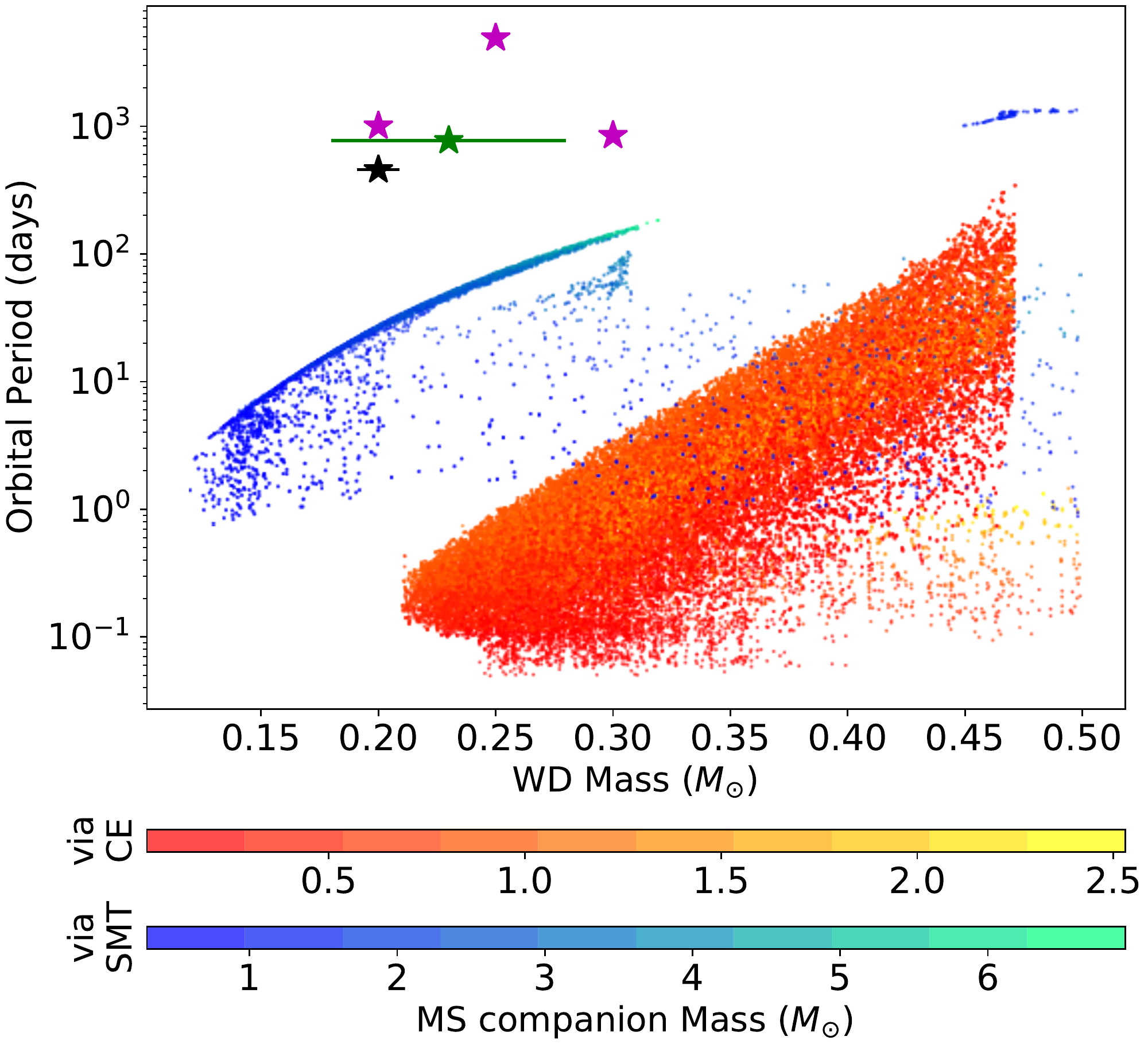}
    \caption{The orbital period and the WD mass for all the EW--MS binaries existing during the $14\,\gyrs$ evolution of a realistic cluster's binary population, evolved using \cosmic. The red-yellow points represent the binaries that have gone through a CE evolution during RLOF (via CE), while the blue-green points represent the ones that only go through SMT (via SMT). The colors along with the color bars denote the masses of their MS companions in $\msun$. The black star denotes the location of the \ab,\ which is at least an order of magnitude wider than any binary in the WD mass range of $0.15-0.25 \msun$. The other interesting wide binaries from \autoref{fig:survey} are also reproduced here. Clearly, the best possible population synthesis of isolated binaries cannot reproduce the properties of the \ab. 
    }
    \label{fig:cosmic_kic}
\end{figure}

In order to verify whether EW--MS binaries similar to the \ab\ can form at all via isolated binary star evolution, we use the population synthesis code, \cosmic\ \citep{Breivik2020}, to evolve $10^6$ zero-age MS binaries up to $14\,\gyrs$. We draw the initial orbital periods from a distribution flat in $\log\porb$ \citep{1983ARA&A..21..343A}. We assume that the initial eccentricities are thermal \citep{Jeans1919}. The \ab\ is observed to have high metallicity, $Z\approx0.05$ (\autoref{tab:ab}). Hence, we assign the highest supported metallicity in \cosmic, $Z=0.03$, for all binaries we simulate. In \autoref{fig:cosmic_kic}, we show the $\porb$ and WD mass for all EW--MS binaries formed at any time within $14\,\gyrs$ in our simulations. As expected, all EW--MS binaries go through a phase of RLOF. The red-yellow and blue-green points represent binaries that have gone through CE and SMT, respectively. The color scale denotes the MS companion's mass. The black star denotes the \ab, which is at least an order of magnitude wider than any binary with a similar WD mass. This bolsters the understanding that although the mass transfer in a binary can produce EW binaries, the $\porb$ of the \ab\ is too long compared to what is expected. In what follows, we consider the possibility of dynamically changing the orbital properties of a typical EW binary, which could form through binary stellar evolution, to those observed in the \ab. 

\section{Analytic considerations for binary-mediated interactions}
\label{sec:theory}
 Throughout the paper, we consider a fiducial EW binary to have the following properties. We use the observed mass of the EW in the \ab\ ($\melmwd/\msun=0.2$). We adopt the mean companion mass ($M_c/\msun=0.76$) and the median orbital period ($\porb/\hrs=5.4$) of the observed EW binaries as the companion mass and the orbital period, respectively \citep{Brown2016}. We further assume that initially this binary is circular and the companion is a regular CO WD \citep[since this is thought to be the most common companion type for the observed EW binaries;][]{Andrews2014,Boffin2015}. 

First, we consider binary-single scattering, the simplest binary-mediated interaction, where the fiducial EW binary is perturbed by a single MS star of mass $\sim1.1\,\msun$, similar to the one in the \ab. The two most relevant quantities determining the outcomes of such interactions are the critical velocity ($v_c$)---defined as the value of the relative velocity at infinity between the binary's center of mass and the single star, $\vinf$, such that the total energy of the three-body system is zero in the barycenter frame of the three bodies \citep{1996ApJ...467..359H}---and the orbital speed ($\vorb$).
For our fiducial EW binary, 
\begin{eqnarray}
\label{eq:vc}
v_c & = & \left( \frac{G \melmwd M_c}{\mu a} \right)^{0.5} \nonumber \\ 
       & = & 140.5 \left( \frac{\mpert}{1.1 \msun}  \right)^{-0.5} \nonumber\\
       & & \times \left( \frac{\mpert}{1.1 \msun}+0.87 \right)^{0.5}\,\kms, 
\end{eqnarray}
and 
\begin{eqnarray}
\label{eq:vorb}
\vorb & = & \left(G \frac{\melmwd + M_c}{a} \right)^{0.5} \nonumber. \\
    & = & 346\,\kms
\end{eqnarray}
Here, $\melmwd$, $M_c$, and $\mpert$ denote the masses of the EW, its companion, and the single-star perturber, respectively. $\mu=M_{\bin}\mpert/M_{\tot}$ is the reduced mass of the three-body system, $M_{\bin}=\melmwd+M_c$ and $M_{\tot}=M_{\bin}+\mpert$. $a$ is the semi-major axis (SMA) of the EW binary and $G$ is the gravitational constant.

\begin{figure}
    \centering
    \epsscale{1.2}
    \plotone{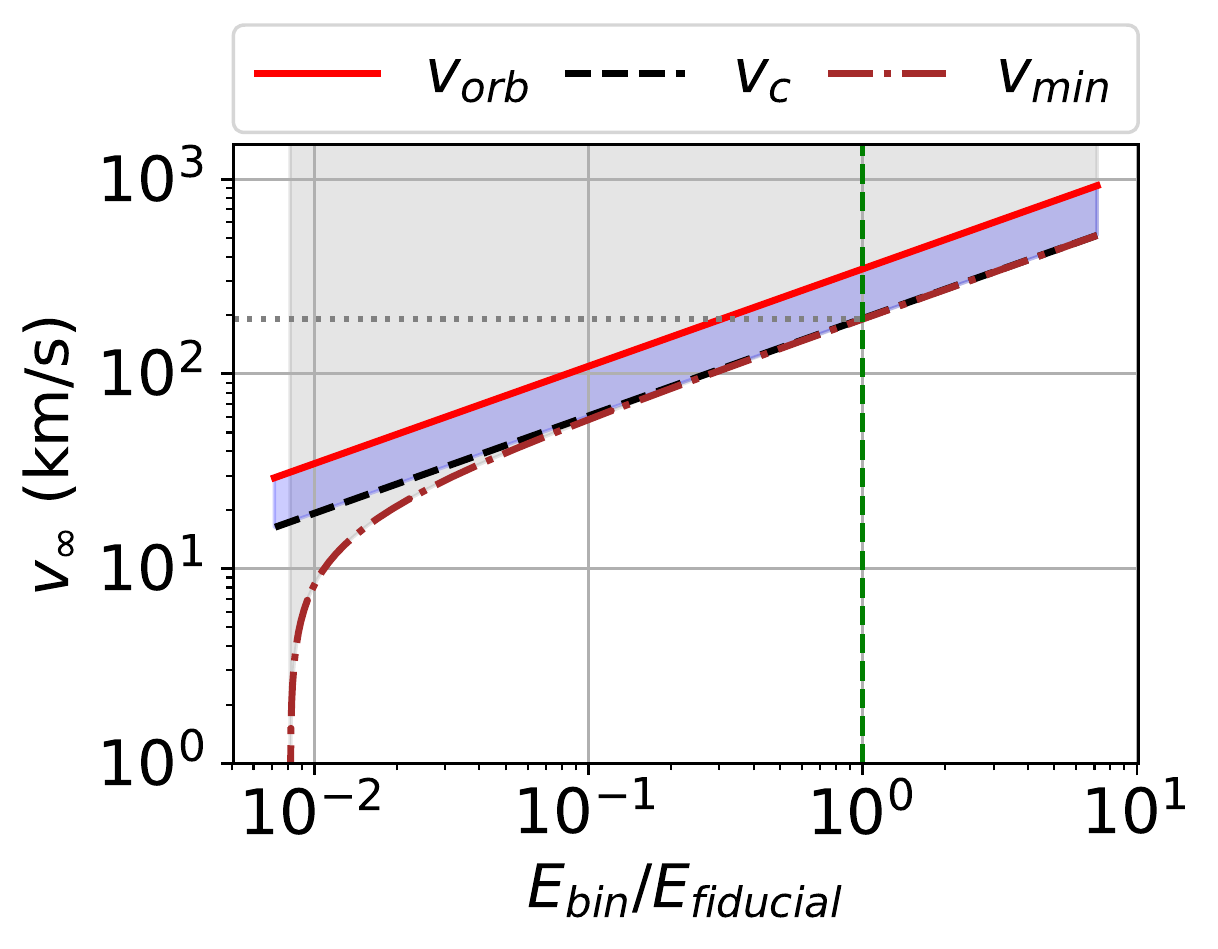}
    \caption{The orbital velocity, $\vorb$, the critical velocity, $v_c$ and the lower limit on $\vinf$ to fulfill the energy requirement, $\vmin$ are plotted for our chosen binary-single system as a function of the binary energy (normalized to the energy of the fiducial binary). The green vertical dashed line represents the energy of our fiducial binary.}
    \label{fig:energy}
\end{figure}

The parameter space of interactions can be divided into characteristically different regions using $v_c$ and $\vorb$ (\autoref{fig:energy}). The interactions in these different regions result in statistically different outcomes \citep{Fregeau2006}. When $\vinf < v_c$, the total energy of the system is less than zero. Consequently, most of the strong encounters would be resonant, i.e., all the three stars would form a temporary bound state and will go through multiple close passages \citep{Heggie2003}. A bound state like this would statistically result in the eventual ejection of the lightest star, which is the EW in our case. Thus, this region does not provide ideal conditions for the assembly of an EW--MS binary. On the other hand, if $\vinf > \vorb$, the interaction time scale would be smaller than the orbital period, due to the fast-moving intruder. An ``impulsive" strong encounter like this would statistically result in ionization \citep{Fregeau2006}, i.e., all three stars would become singles. Thus, we do not expect this region to be ideal for the production of EW binaries. This leaves a strip of parameter space, $v_c<v_{\infty}<\vorb$, where the interactions are not energetic enough to ionize, but are too energetic to be dominantly resonant. 

In addition to the right kind of exchange outcome, we also need enough energy to widen the dynamically formed EW binary orbit. The lower limit for $v_{\infty}$ can be estimated by considering that the entire initial energy of the three-body system is transferred to the final EW--MS binary: 
\begin{eqnarray}
\label{eq:binsingle2}
\vmin & = & \left[G\melmwd\ \frac{M_{\tot}}{M_{\bin} \mpert}\ \left( \frac{M_c}{a_{\ini}} - \frac{\mpert}{a_{\fin}} \right) \right]^{0.5} \nonumber\\
& \approx & 192\,\kms,
\end{eqnarray}
where, $a_\ini$ and $a_\fin$ denote the SMA of the initial fiducial EW binary and the final EW--MS binary, respectively. We obtain the numerical value by using $\mpert/\msun=1.1$ and $a_\fin=1.27\,\au$, similar to the \ab.

All of the above suggests that in order to convert the fiducial EW binary into an EW--MS binary as wide as the \ab\ via binary-single encounters, we require $v_{\infty}>\vmin$ (the gray shaded region in \autoref{fig:energy}) as well as $v_c<v_{\infty}<\vorb$ (the blue shaded region). Thus, the required $v_{\infty}$ is clearly about an order of magnitude higher compared to the typical velocity dispersion of star clusters. 
Thus, while such exchange encounters can be common, especially in stellar clusters where the typical velocity dispersion ($\sim$a few $\kms$) is lower than the $\vorb$ of our fiducial EW binary \citep[e.g.,][]{Fregeau2004}, it is energetically challenging to create an EW--MS binary with an orbit as wide as the \ab\ via binary-single exchange encounters.

Another possibility is that the initial binary itself was an EW--MS binary in a tight orbit within the limiting $\porb$ from the requirement of mass transfer, and binary-single flyby (preservation) encounters may have widened it. Even in this case, we can estimate $\vmin\approx 214\,\kms$ assuming, in this case, $M_c=1.1\,\msun$ in \autoref{eq:binsingle2}, with everything else being kept fixed. 

While these required values for $v_\infty$ are obtained adopting the fiducial binary properties, it is clear how they would change with the SMA. Both $\vorb$ and $v_c$ vary as $a^{-0.5}$ (\autoref{eq:vc} and \autoref{eq:vorb}). For example, instead of adopting the median $\porb$ of the observed EW binaries, if we adopt the maximum $\porb/\days=26$ found in our binary stellar evolution simulations (\autoref{sec:popsynth}), we would find $v_\infty/\kms\gtrsim 35\ (40)$ to create an EW binary as wide as the \ab\ via exchange (preservation), much larger than the observed velocity dispersion in typical star clusters.

These analytic considerations indicate that we must consider binary-binary interactions at the least, since in case of a binary perturber, its binding energy provide an additional source of energy. In the rest of this paper, we consider in detail binary-binary encounters between the fiducial EW binary and a double-MS binary. In particular, we examine whether the fiducial EW binary can be converted into a binary similar to the \ab\ via binary-binary interactions. 

\section{Numerical setup for binary-binary interactions}
\label{sec:method}
We simulate binary-binary interactions using \fewbody, a general purpose small-$N$ dynamics code well suited for scattering experiments \citep{Fregeau2004,Fewbodycode2012}. \fewbody\ employs an eighth-order Runge-Kutta Prince-Dormand integration method, adaptive time step, and global pairwise Kustaanheimo-Stiefel (K-S) regularization \citep{Heggie1974,Mikkola1985}. Throughout the paper, we denote the four objects involved in binary-binary interactions via the indices $0$--$3$. The members are enclosed within ``[ ]" to denote bound systems. The parent members are combined using ``:" to denote collisions. For all our simulations, we use sticky sphere collisions. The initial properties of the binaries and the individual stars and their indices are summarized in \autoref{tab:inprop}. Note that, $[2\ 3]$ denotes the EW--WD binary that initially has the same properties as the observationally guided properties of the fiducial EW binary described in \autoref{sec:theory}. The properties of the regular WD are not very important, since they are lost from the system in our intended outcomes.     

\begin{deluxetable}{ccccc}
\tablecolumns{5}
\tabletypesize{\footnotesize}
\tablewidth{0pt}
\tablecaption{Initial properties of the stars and binaries.\label{tab:inprop}}
\tablehead{
	\colhead{Index} & \colhead{Object} & \multicolumn{3}{c}{Properties} \\
    \cline{3-5}
    \colhead{} & \colhead{} & \colhead{Mass} & \colhead{SMA} & \colhead{$e$}\\
    \cline{3-4}
    \colhead{} & \colhead{} & \colhead{$(M_\odot)$}& \colhead{$(\au)$}& \colhead{} 
    }
\startdata
 0 & MS & 1.1 & \nodata & \nodata \\ 
 1 & MS & 1.0 & \nodata & \nodata \\
 2 & EW & 0.2 & \nodata & \nodata \\
 3 & WD & 0.76 & \nodata & \nodata \\ 
 \hline\\
 {[}0 1{]} & MS--MS & 2.1 & Uniform in log(\rm{SMA}) & Thermal \\
 {[}2 3{]} & EW--WD & 0.96 & 0.007 & 0 \\
\enddata
\end{deluxetable}
We choose the members of the MS binary (denoted by $[0\ 1]$) to have masses similar to the MS star in the \ab. 
The initial orbital properties of $[0\ 1]$ are guided by what is expected in a typical star cluster. For example, we draw the initial SMA from a distribution flat in $\log\ \rm{SMA}$ \citep{1983ARA&A..21..343A} between $5\times(R_0+R_1)$ ($R_i$ denotes the radius of the $i$th star) to the hard-soft boundary ($\ahs$), given a velocity dispersion $v_\sigma$. We treat $v_\sigma$ as a parameter and vary it over a wide range relevant for star clusters. The initial eccentricities are drawn from a thermal distribution \citep[$f(e) = 2e$;][]{Jeans1919}. The $[0\ 1]$ and $[2\ 3]$ binaries approach each other along hyperbolic trajectories with velocity at asymptotic infinity, $v_{\infty}$ drawn from a Maxwellian distribution corresponding to a line-of-sight rms velocity of $v_\sigma$, truncated at the escape speed, $\vesc=2v_\sigma$. For each encounter, we randomize the orientations of the binaries assuming isotropy and randomize the phases in the allowed range. 

Note that in our setup, the initial SMA of $[2\ 3]$, $a_{[2\ 3]}$, is fixed to the fiducial value adopted to be the median of the observed EW binaries, whereas the initial $a_{[0\ 1]}$ is varied in the full allowed range. The former is a reasonable simplification that allows us to somewhat reduce the initial parameter space. We will see later (\autoref{sec:results}) that the properties of the final wide EW--MS binaries are determined by $a_{[0\ 1]}$. Moreover, since any interaction resulting in a wide EW--MS binary must have $a_{[0\ 1]}\gg a_{[2\ 3]}$, the interaction cross sections are also determined by $a_{[0\ 1]}$. Thus, none of our main results are expected to depend on the adopted $a_{[2\ 3]}$ within the observed range for EW binaries.

We consider nine values of $v_\sigma$ between $1$ and $40\,\kms$. For each $v_\sigma$, we perform at least $10^4$ binary-binary scattering experiments with impact parameter $b$ between 0 and $\bmax$, where $\bmax$ is the value of the impact parameter such that the distance of closest approach $\rperi=2\times(a_{[2\ 3]}+0.2\,\au)$ along the hyperbolic trajectory, with $\vinf$ = $v_\sigma$ taking into account gravitational focusing. The impact parameter is chosen successively from adjacent larger annular regions, starting with $b=0$ for successive interactions. We perform one scattering experiment with an impact parameter randomly chosen between $b$ and $b+\delta b$, where $\delta b=\bmax/10^4$. Once the interaction is over, we analyze the outcome and add the annular area within $b$ and $b+\delta b$ to the cross section corresponding to that particular outcome. We repeat this process up to $b=\bmax$. This way, the code estimates the cross sections of all the possible outcomes, capturing even their dependence on the impact parameter. We ensure that the final $\bmax$ is sufficiently large such that we do not miss any of our intended outcomes, i.e., final binaries [0 2] or [1 2]. We take note of the last impact parameter ($\blast$) where anything other than weak flyby and pathological collision happens and update $\bmax$ to $2\blast$ if $2\blast>\bmax$.\footnote{Depending on the choice of the SMA and $e$, the members of the MS binary [0 1] may collide with each other even without any dynamics. We call these pathological collisions. We identify them as systems where MS stars 0 and 1 collide without any precollision change in the initial SMA of the [0 1] binary.} We continue scattering experiments up to the updated $\bmax$ using the same $\delta b$, thus also increasing the total number of scattering experiments performed. We calculate cross sections for all outcomes except weak flyby and pathological collisions. This technique to calculate cross sections for different outcomes was introduced in \citet{Fregeau2006}. Our implementation closely follows that in the code {\tt sigma\_binsingle}, a part of \fewbody's numerical toolkit, modified for binary-binary interactions and our specific problem. 

For each $v_\sigma$, we repeat this exercise $320$ times to take into account statistical fluctuations in the cross sections of the different outcomes of interest. This allows us to estimate the statistical error bars on the cross sections for each outcome category. In total, we simulate $\sim7\times10^{11}$ binary-binary scattering experiments. Since our primary goal is to investigate the possible production of wide EW binaries similar to the \ab, we  focus on outcomes containing $[0\ 2]$ or $[1\ 2]$ binaries. We combine all other outcomes (except weak flyby encounters and pathological collisions between $[0\ 1]$ binary members) in our cross section calculation. We further categorize the outcomes that resulted in an EW--MS binary ([0-or-1 2]) in three categories:
\begin{itemize}
\item Wide binaries: these are EW--MS binaries with an SMA above the SMT boundary of \citet{Lin2011}.
\item Ejected binaries: these are EW--MS binaries where the recoil kicks from the interaction are sufficient to eject them from the host cluster. We assume that the escape speed $\vesc=2v_\sigma$. 
\item Ejected wide binaries: these are EW--MS binaries that satisfy both of the above conditions.
\end{itemize}

\section{Results}
\label{sec:results}

\subsection{A viable formation pathway}
\label{sec:viability}
\begin{figure}
    \centering
    \epsscale{1.2}
    \plotone{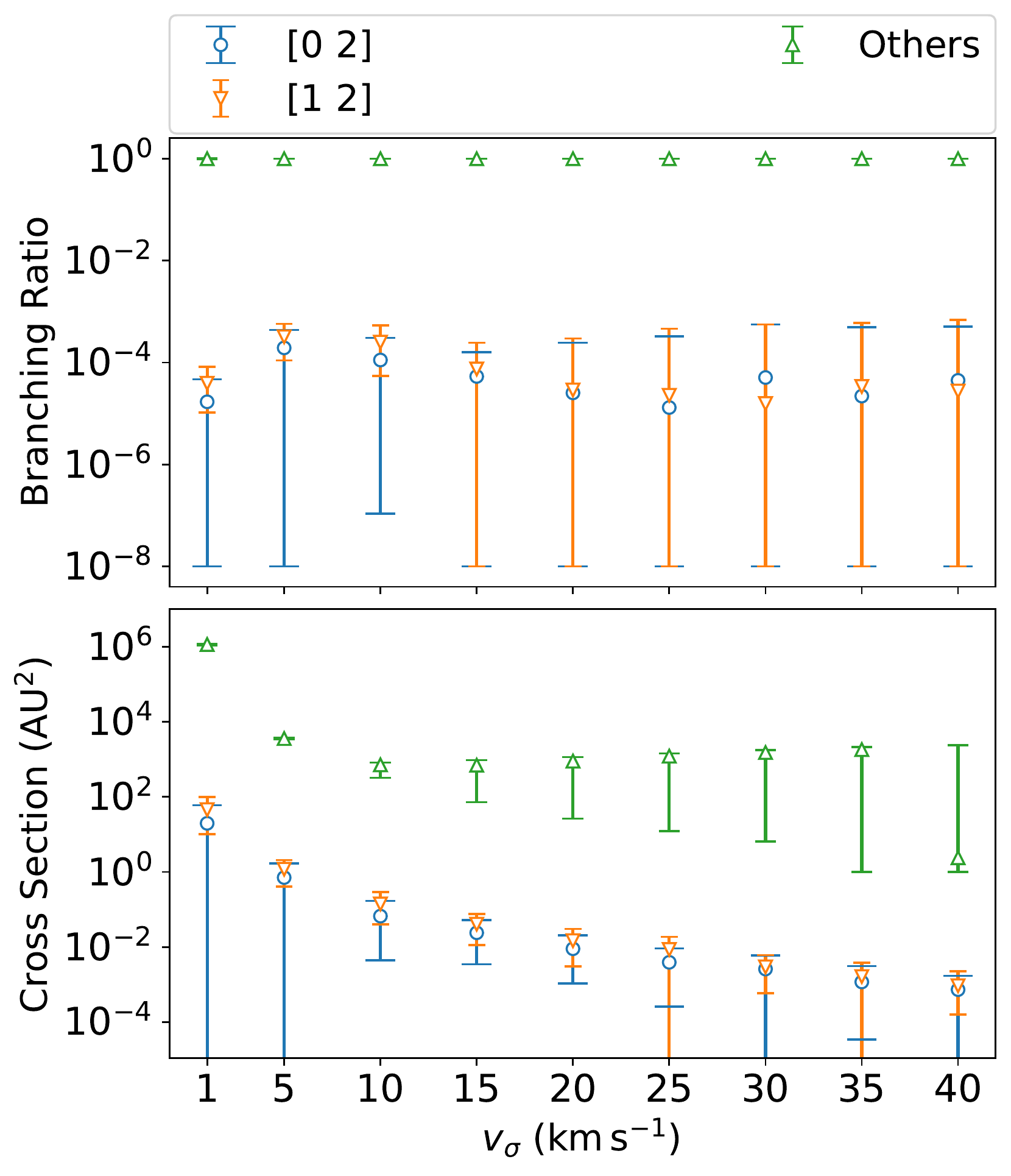}
    \caption{Branching ratios (top) and cross sections (bottom) computed from our simulations. `[0 2]' and `[1 2]' combine every encounter that results in a [0 2] or [1 2] binary, respectively. `Others' combines the encounters that do not form an EW--MS ([0 2] or [1 2]) binary. We calculate the cross sections 320 times for each setup by varying the random seeds for every value of $v_{\sigma}$ resulting in a distribution of 320 different values of cross sections and branching ratios. The markers represent the mode of the distribution. The errors denote $1\sigma$ around the mode. We find that the cross sections for the `Others' outcome are bimodal for $v_{\sigma}\gtrsim15\,\kms$. At $v_{\sigma}=40\,\kms$, the mode shifts from the higher peak to the lower peak. This causes a visible shift of the mode, apparently breaking the trend at $v_{\sigma}=40\,\kms$. Note that the $1\sigma$ ranges are more meaningful here. 
    }
    \label{fig:csbrl}
\end{figure}

In \autoref{fig:csbrl}, we show the cross sections, $\sigma$, and branching ratios for the encounters that result in an EW--MS binary (`[0 2]' and `[1 2]'), along with the ones that do not (`others'). The branching ratio for a particular outcome is defined as the ratio of the cross section of that outcome to the sum of the cross sections of all outcomes. We find that the branching ratios for `[0 2]' and `[1 2]' are small ($\lesssim10^{-3}$) across the range of $v_{\sigma}$. Furthermore, the branching ratios for different outcomes are largely unaffected by $v_{\sigma}$, while the cross sections decrease with increasing $v_{\sigma}$. This is primarily due to the inverse square dependence of $\vinf$ in the gravitational focusing term. In other words, a fast-moving intruder would need to approach at a smaller impact parameter in order to have the same $\rperi$ to the target as a slower intruder. Moreover, even if two interacting binaries have a fixed $\rperi$, the outcome of the interaction may still depend on $\vinf$ due to the dynamics of the encounter. The net result of all these effects is captured in \autoref{fig:csbrl}.

\begin{figure*}
    \centering
    \epsscale{1.15}
    \plotone{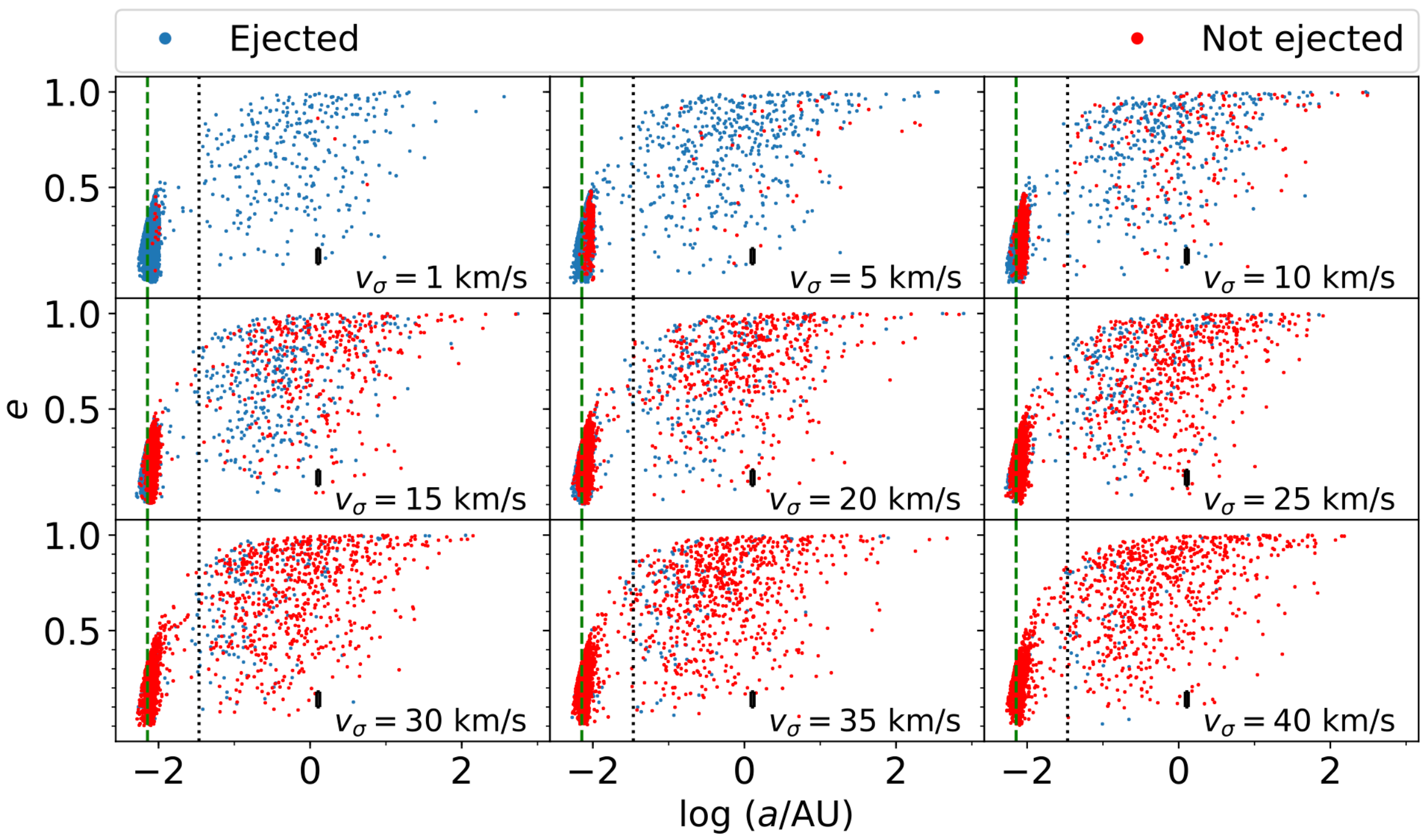}
    \caption{SMA vs eccentricity scatter plots for all the EW--MS binaries, [0 2] and [1 2], that form in our simulations for different velocity dispersions. The black rectangles mark the location of the observed \ab{} within twice the $1\sigma$ error. The green dashed vertical line marks the median SMA for EW binaries from the survey of \citet{Brown2016}. This value is also the fixed initial SMA of [2 3] in our simulations. The black dotted vertical line marks the \citet{Lin2011} SMT boundary. The blue dots represent the binaries that receive enough kick during the encounter to get ejected from the host cluster, while the red ones represent those that do not. }
    \label{fig:ae}
\end{figure*}

In \autoref{fig:ae}, we show the orbital properties of all EW--MS binaries formed in our simulations. The different panels show the results for different adopted $v_\sigma$ values. The vertical lines denote the initial SMA of the $[2\ 3]$ binary (green dashed) and the SMT limit \citep[black dotted;][]{Lin2011}. The blue and red dots denote binaries that would be ejected from and retained in the cluster, based on the $\vrecoil$ given by the simulations and the expected $\vesc$. The location of the \ab\ with twice the $1\sigma$ errors is marked by the black rectangles. While a large number of EW--MS binaries are formed with SMAs very similar to that of the initial [2 3] binary, a significant fraction ($18\%$) attain SMAs larger than the limiting value for the onset of SMT \citep{Lin2011}. Although for all adopted $v_\sigma$ the wide EW--MS binaries show a combination of ejected and retained systems, there is a clear trend. Lower-$v_\sigma$ models show a relatively higher fraction of wide ejected EW--MS binaries. This is simply because a lower $v_\sigma=0.5\vesc$ makes it easier for the binary to escape the cluster, due to recoil. Interestingly, the ejected and retained systems do not show much difference in the parameter space they populate in the $a-e$ plane. All wide EW--MS binaries show a wide range in eccentricities, including the low observed eccentricity ($e<0.14$) of the \ab{} (\autoref{tab:ab}). Thus, it is possible to widen a typical EW binary orbit via binary-binary encounters to high SMAs similar to the observed \ab.

\subsection{Formation channels}
\begin{figure}
    \centering
    \epsscale{1.2}
    \plotone{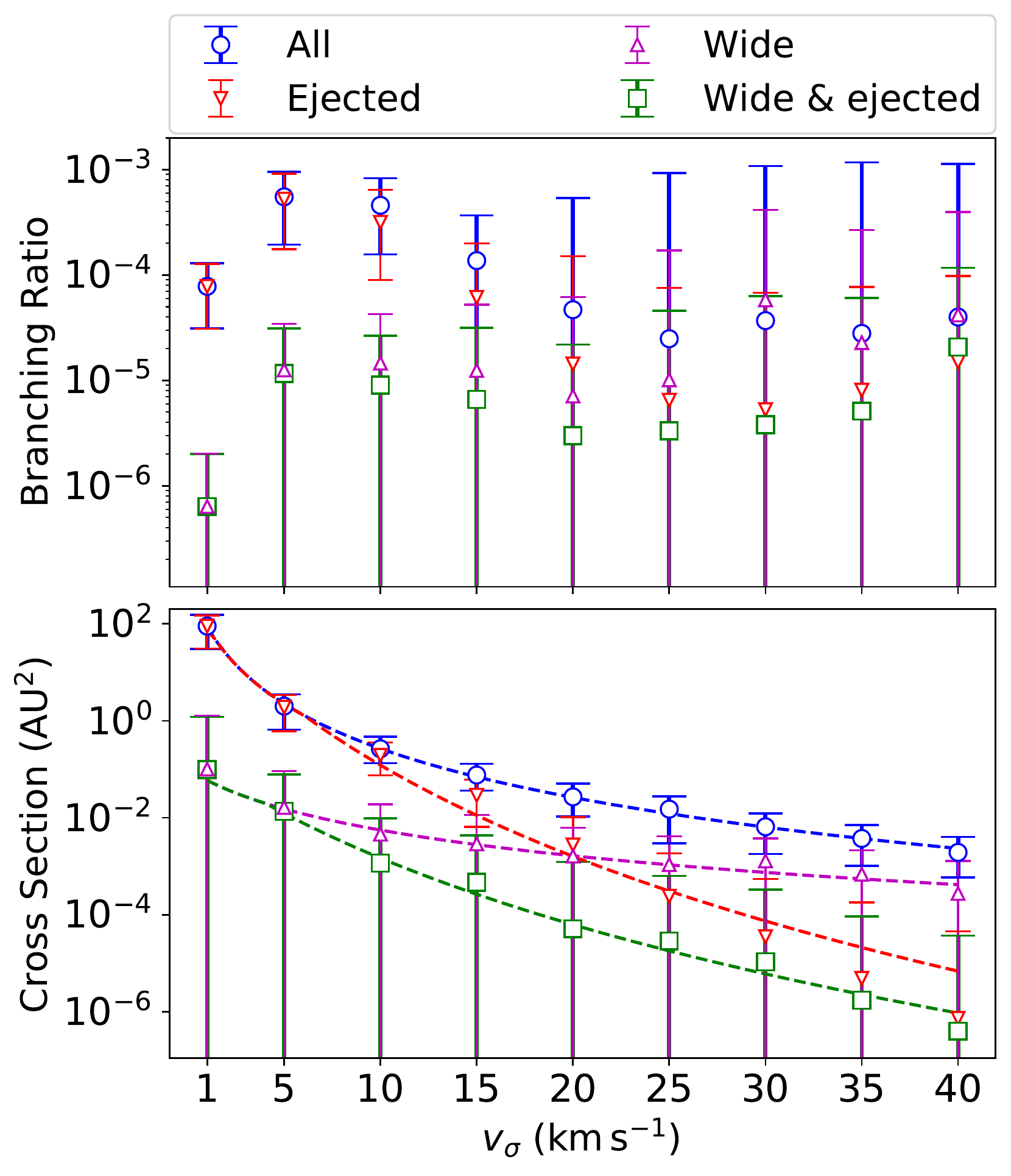}
    \caption{The same as \autoref{fig:csbrl}, except here we only analyze the successful encounters, i.e., the ones resulting in an EW--MS binary. `All' combines `[0 2]' and `[1 2]' from \autoref{fig:csbrl}. `Wide' requires that the binary's SMA is greater than the value obtained by the \citet{Lin2011} expression for a WD of mass $0.2\,\msun$. The idea is to define a population that cannot be formed through the SMT channel. `Ejected' requires the binary formed to receive enough kick from the encounter to get ejected from its host cluster. `Wide and ejected' are encounters satisfying the constraints of both `ejected' and `wide.'}
    \label{fig:csbrr}
\end{figure}
Here, we focus on only the encounters that form EW--MS binaries. First, we categorize these encounters based on their SMA and recoil velocity. 
 In \autoref{fig:csbrr}, we show the cross sections, $\sigma$, and branching ratios for all EW--MS binaries ($\sigmaall$; blue circles), those that are in orbits wider than the \citet{Lin2011} boundary ($\sigmawide$; purple upright triangles), those that are expected to be ejected from the host cluster ($\sigmaejected$; red inverted triangles), and those that are both in wide orbits and are expected to be ejected ($\sigmawideejected$; green squares), as a function of $v_{\sigma}$.  We fit a power law of the form, $\sigma=j(v_{\sigma}+k)^{-l}$ to obtain the cross sections for these various outcomes as a function of $v_\sigma$, where $j$, $k$, and $l$ are the fitting parameters. The best-fit values are given by:
 \begin{eqnarray}
 \label{eq:cseq}
 \frac{\sigmaall}{\au^2} &=& 1.92\times10^3\: \left(\frac{v_{\sigma}}{\kms} + 1.33\right)^{-3.66} \nonumber \\
 \frac{\sigmaejected}{\au^2} &=& 9.00\times10^{12}\: \left(\frac{v_{\sigma}}{\kms} + 9.96\right)^{-10.67} \nonumber \\
 \frac{\sigmawide}{\au^2} &=& 2.05\: \left(\frac{v_{\sigma}}{\kms} + 3.85\right)^{-2.25} \nonumber \\
 \frac{\sigmawideejected}{\au^2} &=& 4.73\times10^9\: \left(\frac{v_{\sigma}}{\kms} + 13.96\right)^{-9.07}.
 \end{eqnarray}
In addition, we ensure that the cross section for a subset outcome is never larger than its superset.\footnote{ Unrestricted power-law fits may erroneously create situations where the subset cross section, as predicted by the fit, is marginally larger than its superset, although the measured cross sections are equal. This of course is an artifact. We eliminate this by invoking $\sigmaall\geq$ both $\sigmaejected$ and $\sigmawide$, and $\sigmawideejected\leq$ both $\sigmawide$ and $\sigmaejected$ for all $v_\sigma$.} 

The cross section for the creation of EW--MS binaries decreases as $v_{\sigma}$ increases, as discussed in \autoref{sec:viability}. Moreover, we find that $\sigmaejected$ decreases faster than $\sigmawide$. This is simply because $\vesc = 2\,v_{\sigma}$ and as $v_\sigma$ increases, $\vrecoil$ must proportionally increase for a successful ejection.
\begin{figure*}
    \centering
    \epsscale{1.15}
    \plotone{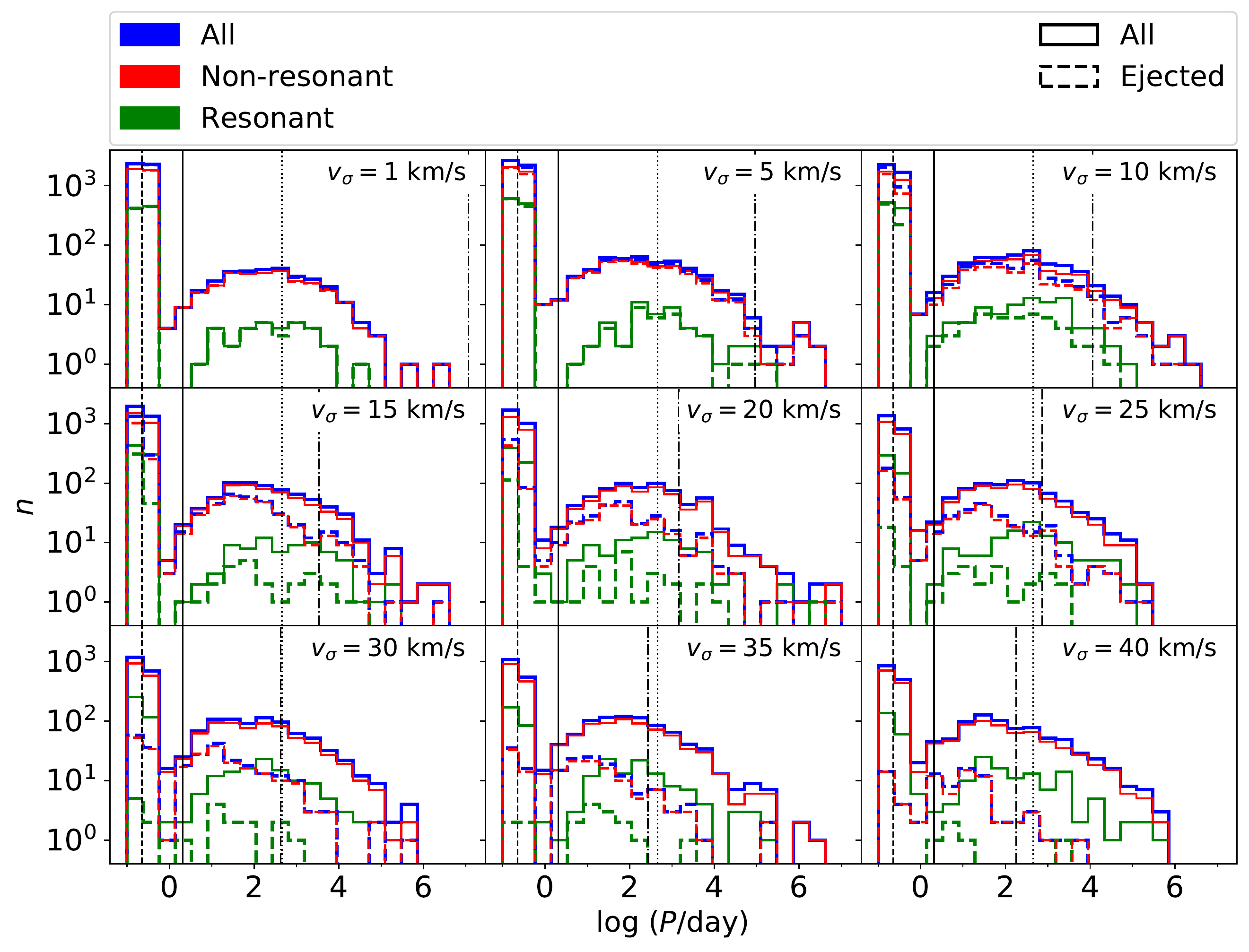}
    \caption{Histograms for the orbital periods (days) of the EW--MS binaries (i.e., [0 2] or [1 2]) formed in our simulations. The different line styles of the histograms categorize the binaries based on their recoil velocity, $\vrecoil$. The solid histograms include all the binaries irrespective of the value of $\vrecoil$, while the dashed histograms only includes those binaries that receive enough $\vrecoil$ during the encounter to get ejected from their host cluster. The different colors tell us about the nature of the encounters. The red and green histograms include all the binaries that formed via nonresonant and resonant encounters, respectively, while the blue histograms combine the two. The vertical lines mark the important period scales in our analysis. The dashed line shows the median orbital period from the \citet{Brown2016} survey. This is also the fixed initial orbital period of the EW--WD binary (i.e., [2 3]) in our simulations. The solid line marks the orbital period for a binary containing a $0.2\,\msun$ EW that can be formed via the SMT channel, according to the expression given in \citet{Lin2011}. The dotted line shows the observed orbital period of the \ab{}. Finally, the dashed-dotted line mark the hard-soft boundary for the final EW--MS binary for each panel.}
    \label{fig:period1}
\end{figure*}

In \autoref{fig:period1}, we show the distribution of orbital periods for all (solid) final [0 2] and [1 2] binaries, and those that are ejected (dashed). In addition, we show the subsets of [0 2] and [1 2] binaries created via resonant (green) and nonresonant (red) encounters. Several trends emerge. As already seen in \autoref{fig:csbrr}, the fraction of ejected systems decreases with increasing $v_\sigma$. In addition, for all $v_\sigma$, nonresonant encounters dominate the production of [0 2] and [1 2] binaries. This is expected, since resonant encounters typically push systems closer to energy equipartition. As a result, the least massive member, the EW in our case, is typically ejected. Since our desired binary contains the EW, nonresonant encounters dominate their production. 

We find that the distribution for orbital periods is bimodal for all $v_\sigma$. The narrow and dominant peak is near the [2 3] binary's initial orbital period (the vertical dashed line), which is also the median orbital period for the known EW binaries \citep{Brown2016}. Most systems around this peak are within the SMT limit (the vertical solid line) given in \citet{Lin2011}. We name the binaries in this part of the distribution the ``tight'' population. The second peak is around $P\sim$ few $10^2\,\days$. The distribution around this peak is quite broad and almost entirely spread over orbital periods above the limit for SMT. We denote this population as the ``wide" population. Between the tight and wide populations there is a clear separation. It is interesting that the tight population shows orbital periods very similar to the observed typical EW binaries \citep{Brown2016}, with separations lower than the SMT limit. By contrast, the wide population peaks very near the orbital period of the observed unusually wide EW binary of interest, the \ab\ \citep[the vertical dotted line;][]{Masuda2019}. This bimodal $\porb$ distribution can be understood based on the energetics of the binary-binary interactions that create EW--MS binaries. We delve into this in detail later (see \autoref{sec:nrde}).

While a different adopted $v_\sigma$ changes the relative contributions from the ejected and retained systems, the bimodality in orbital periods, the locations of the peaks, and the separation between the tight and wide populations do not significantly depend on $v_\sigma$. Nevertheless, note that the tail of the distribution at very high $\porb$ values is unphysical, especially for high-$v_\sigma$ cases where most of the EW--MS binaries are \emph{not} ejected. In our setup, we only consider one scattering event, and do not consider a further chance of scattering. In reality, inside a real cluster, the wide binaries can interact repeatedly until ejected or broken. As a result, the binaries that are wider than the hard-soft boundary would be broken or the members would change due to exchange encounters. Hence, we also show the hard-soft boundary in each panel (the vertical dashed-dotted line), which essentially denotes that unless they are ejected from the host cluster, the EW--MS binaries wider than this limit may not be safe.  

\begin{figure}
    \centering
    \epsscale{1.2}
    \plotone{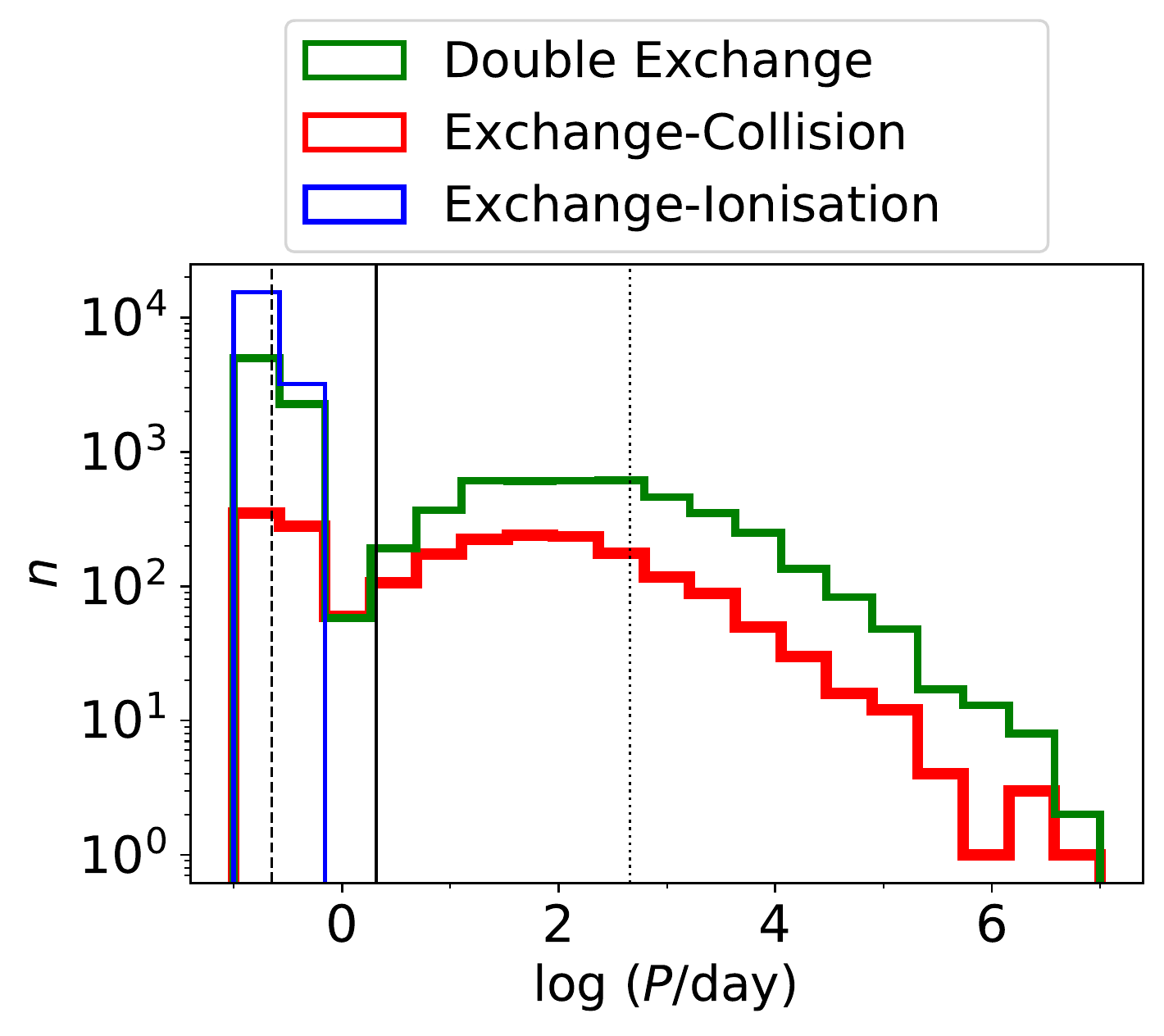}
    \caption{The same as \autoref{fig:period1}, except here we categorize the EW--MS binaries based on the three possible formation channels, depending on the configuration of the remaining two stars taking part in the encounter, i.e., either 0 or 1 and 3 (see the text). `Double exchange,' `exchange-collision,' and `exchange-ionisation' represent the outcomes where the binary [0 3] or [1 3] forms, 3 collides with 0 or 1, and 3 and 0 or 1 are singles, respectively. We also combine the outcomes from all $v_{\sigma}$ simulations into a single panel. Overall, double exchange contributes the most to the production of wide EW--MS binaries.}
    \label{fig:period2}
\end{figure}
\autoref{fig:period2} is very similar to \autoref{fig:period1}, but here we divide the population of [0 2] ([1 2]) binaries depending on the configuration of the other two objects, 1 and 3 (0 and 3). There are three possibilities: 
\begin{itemize}
\item Exchange-ionization: An exchange creates the EW--MS binary, and the remaining stars are single: $[0\ 1]+[2\ 3]\rightarrow[0\ 2]+1+3$ or $[0\ 1]+[2\ 3]\rightarrow[1\ 2]+0+3$.
\item Exchange-collision: An exchange creates the EW--MS binary, and the remaining stars collide with each other: $[0\ 1]+[2\ 3]\rightarrow[0\ 2]+1:3$ or $[0\ 1]+[2\ 3]\rightarrow[1\ 2]+0:3$.
\item Double exchange: Member swap creates an EW--MS and a WD--MS binary: $[0\ 1]+[2\ 3]\rightarrow[0\ 2]+[1\ 3]$ or $[0\ 1]+[2\ 3]\rightarrow[1\ 2]+[0\ 3]$.
\end{itemize}
Here, we combine the results from all $v_{\sigma}$ simulations into a single panel, since we find little difference as a function of $v_\sigma$. We find that, overall, double exchange is the dominant channel for the formation of EW--MS binaries. While exchange-ionization is the dominant channel in the tight EW--MS population for lower $v_{\sigma}$ ($\lesssim25\,\kms$; not shown in the figure), it does not contribute at all to the wide EW--MS population. The shape of the $\porb$ distribution of EW--MS binaries originating from the exchange-collision channel is very similar to that originating from double exchange (more on this in \autoref{sec:nrde}).

\begin{figure}
    \centering
    \epsscale{1.2}
    \plotone{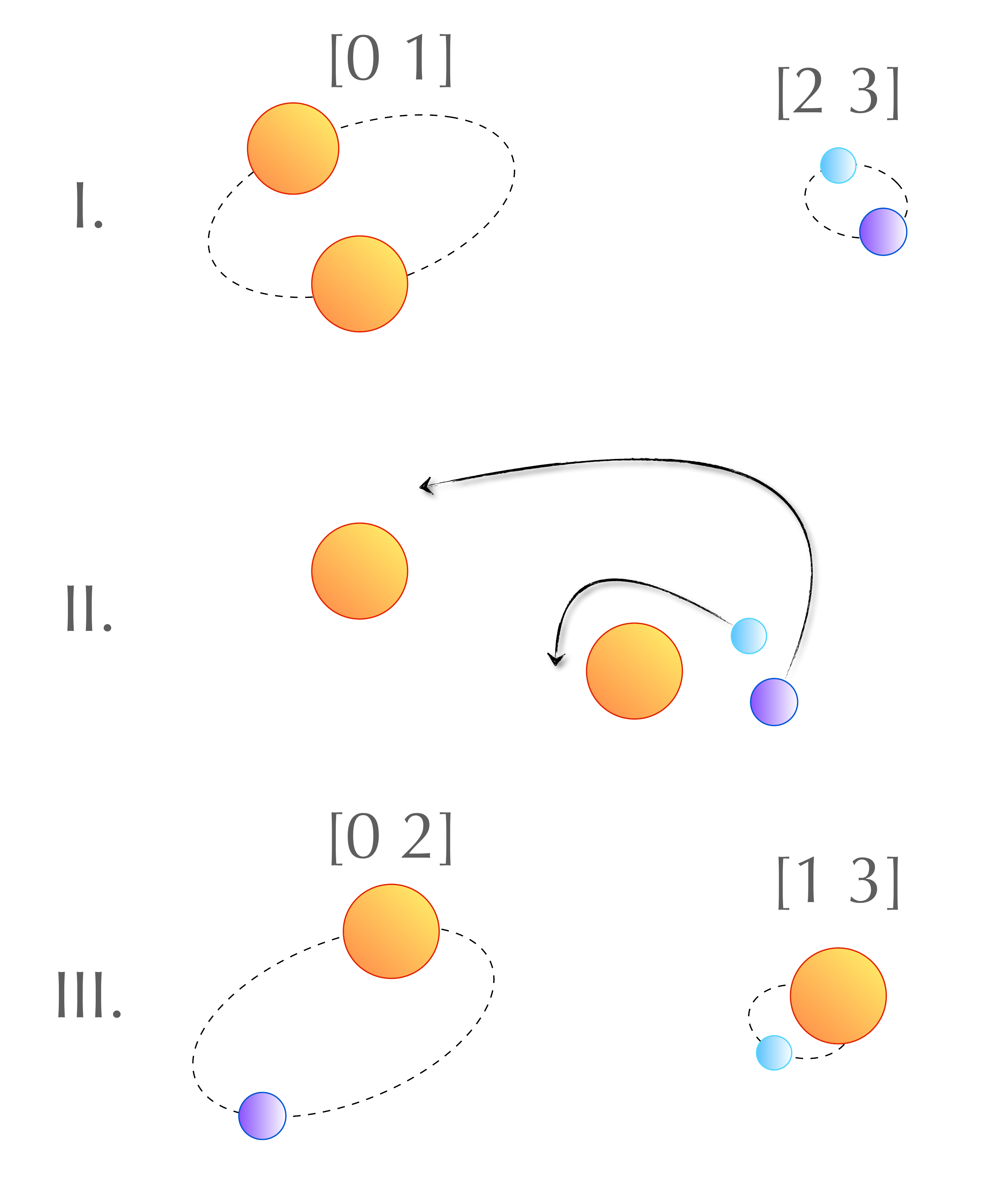}
    \caption{A cartoon depiction of the process of nonresonant double exchange. MS stars are shown as big orange circles, canonical mass WDs as small sky blue circles, and EWs as medium purple circles. \emph{Part I:} the initial configuration of [0 1] and [2 3]. \emph{Part II:} when one of the MS stars gets close enough to [2 3], it exchanges with one of its members, which, in turn, goes into a wider binary with the remaining MS star. Here, the EW is the one that forms the wider binary, [0 2]. \emph{Part III:} the resulting configuration of a wide [0 2] with a tight [1 3].}
    \label{fig:cartoon}
\end{figure}

\subsection{Nonresonant double exchange}
\label{sec:nrde}

Combining \autoref{fig:period1} and \autoref{fig:period2}, we can clearly see that the formation of EW--MS binaries is dominated by nonresonant double-exchange encounters for all the $v_\sigma$ we have considered. The bimodality in the $\porb$ distribution is a direct consequence of the nonresonant double-exchange channel. In \autoref{fig:cartoon}, we show a schematic diagram of the double-exchange channel. The two interacting binaries in our simulations almost always have significantly disparate SMAs: [2 3] is significantly tighter than [0 1]. Consequently, [2 3] would behave essentially like a single object for most interactions. The binary nature of [2 3] would come into play only when one of the two MS stars in [0 1] comes sufficiently close ($\sim$ SMA of $[2\ 3]$) to [2 3]'s center of mass during the interaction. This leaves the remaining MS star of [0 1] far away ($>>$ SMA of $[2\ 3]$) from these three. There are two possibilities in a nonresonant double exchange. The closer MS star could exchange with either 2 or 3 in the [2 3] binary. The exchanged 2 or 3 can subsequently form a binary with the MS star of [0 1] farther away. If the regular WD (star 3) is the one that gets exchanged, it would leave behind an EW--MS binary with orbital period close to that of [2 3]. These are the EW--MS binaries in the tight population. On the other hand, if the EW (star 2) is the one that gets exchanged, it would go on to form a wide EW--MS binary with an SMA close to that of [0 1]. These belong to the wide population. The relative importances of the wide and tight populations are dependent essentially on the mass ratio of the [2 3] binary.  

\begin{figure}
    \centering
    \epsscale{1.2}
    \plotone{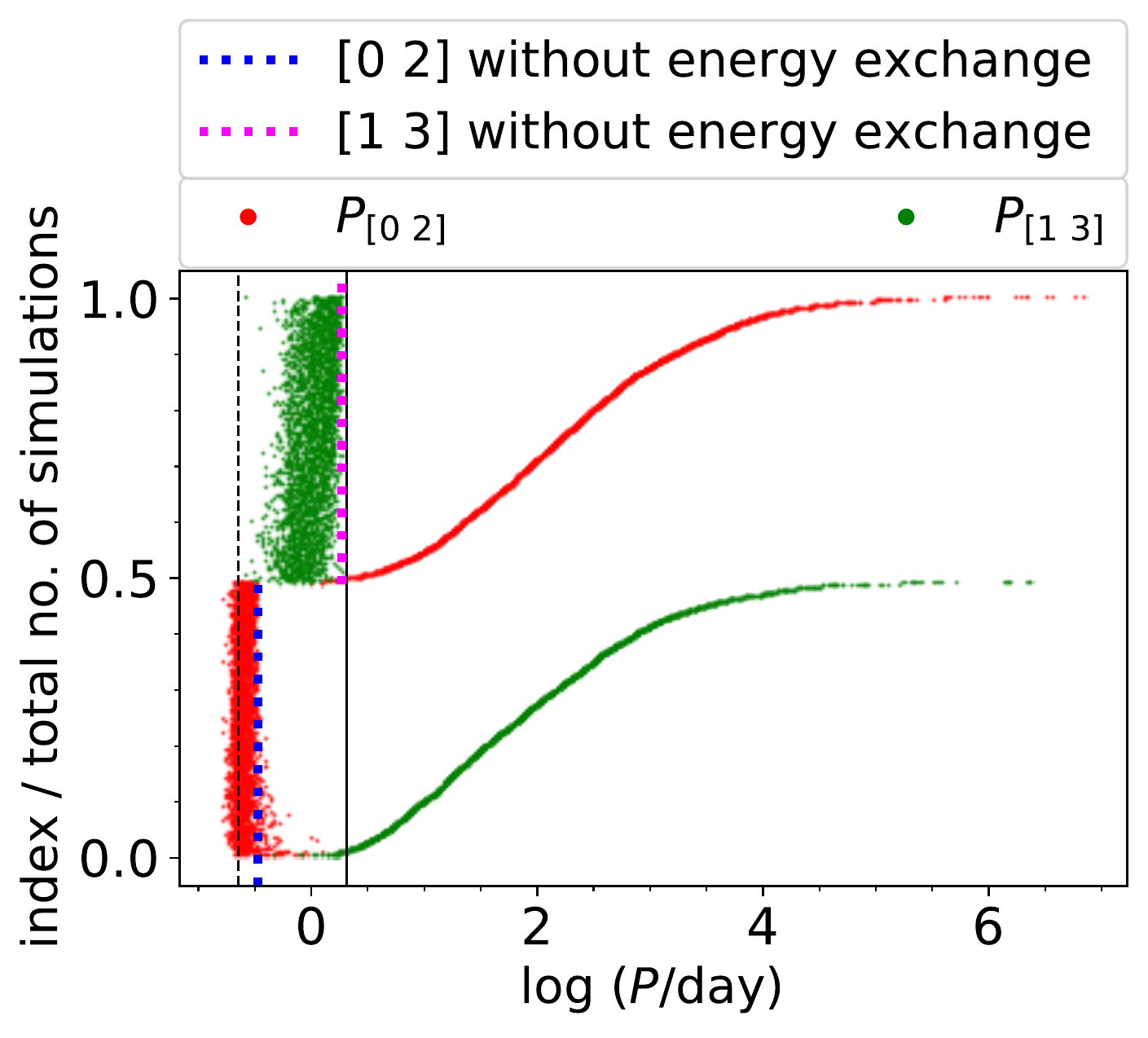}
    \caption{The orbital periods of the [0 2] and [1 3] binaries formed in our simulations, over all values of $v_{\sigma}$. We only include the encounters that resulted in this particular binary-binary outcome of [0 2] [1 3]. A pair of binaries that formed in a particular encounter are plotted with the same $y$-coordinate, while the $x$-coordinate shows their orbital period. The red (green) markers represent the EW--MS binary, [0 2]'s orbital period (the WD--MS binary, [1 3]'s orbital period). The encounters that result in a tighter EW--MS binary compared to the WD--MS binary are grouped together and come first along the $y$-axis, followed by the encounters in which the EW--MS binary is wider. Furthermore, among these two groups, the $y$-coordinate is sorted according to the wide binary's orbital period. This figure confirms the double-exchange scenario, as a tight binary almost always forms with an accompanying wide binary. We also mark the value of the orbital period for the tighter binary if it exactly inherits [2 3]'s initial energy. The dotted blue line marks this value for a tight [0 2], while the magenta line is for a tight [1 3]. As these binaries' orbital periods are generally shorter than the respective zero energy exchange mark, it shows that the exchanged member from  [2 3] takes away more energy than what is brought in by the MS star. The black dashed and solid lines are the same as in \autoref{fig:period1}.}
    \label{fig:nrde1}
\end{figure}
\begin{figure}
    \centering
    \epsscale{1.2}
    \plotone{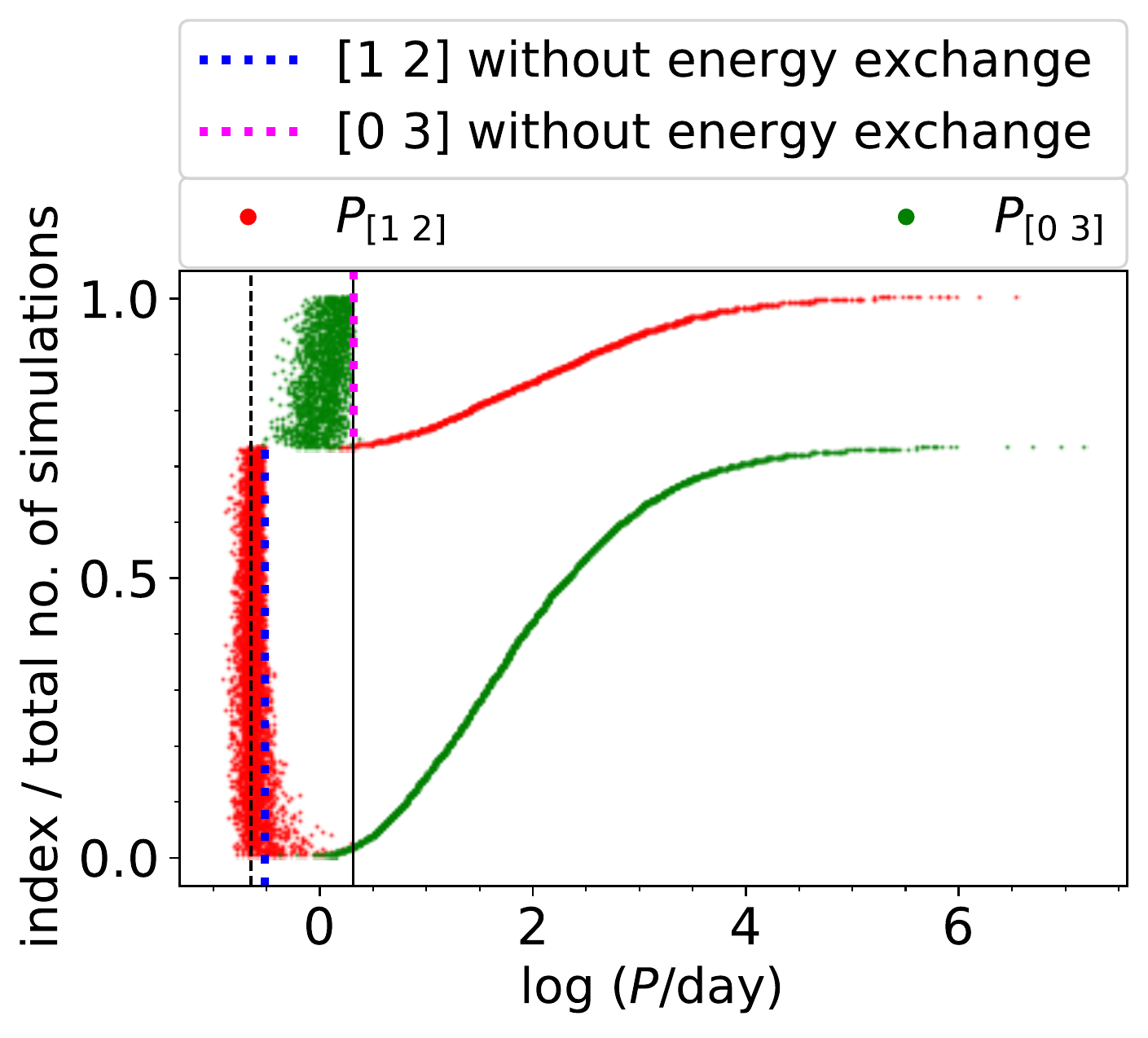}
    \caption{The same as \autoref{fig:nrde1}, except the encounters included here are the ones that resulted in a binary-binary configuration of [0 3] [1 2].}
    \label{fig:nrde2}
\end{figure}
Nonresonant double exchange is essentially a simple swap of the companions with a small amount of energy exchange. We show the pairwise orbital periods of the final binaries formed in all double-exchange encounters leading to either [0 2] (\autoref{fig:nrde1}) or [1 2] (\autoref{fig:nrde2}) EW--MS binaries. These figures include outcomes from all $v_{\sigma}$ simulations. In almost all cases, the two final binaries, in pairs, appear on the two opposite sides of the SMT limit \citep{Lin2011}. The red (green) points denote the orbital periods of the EW--MS (WD--MS) binaries. We find that whether the EW--MS or the WD--MS binary belongs to the wide population does not strongly depend on $v_\sigma$. The EW--MS binary consisting of the $1.1\,\msun$ MS star ([0 2]) has roughly equal probability of belonging to the wide or the tight population (\autoref{fig:nrde1}). In contrast, the outcomes where the EW pairs with the $1\,\msun$ MS star ([1 2]) show a somewhat higher preference of belonging to the tight population ($73\%$; \autoref{fig:nrde2}). This is likely because of the small difference we have imposed on the masses of the two MS stars in our numerical setup. The final tight binary is almost always more bound compared to the initial tight binary ($[2\ 3]$), as expected from Heggie's law \citep{Heggie1974}. Nevertheless, since a more massive MS star exchanges into the initial EW--WD binary in this scenario, the orbital period of the tight binary typically increases. We find that the orbital period of the final tight binary almost always lies between the initial orbital period of $[2\ 3]$ and the orbital period of the final binary corresponding to the initial binding energy of $[2\ 3]$. As a result, the nonresonant double-exchange formation scenario can easily create a wide EW--MS binary that roughly inherits the orbital energy of the wide initial MS--MS binary ($[0\ 1]$), which can widen further by the hardening of the tight binary. 

The relative difference between the widths of the two peaks observed in the $\porb$ distribution of the EW--MS binaries can now be understood by considering the energetics of the dominant double-exchange formation channel. In any dynamical encounter, energy exchange during scattering creates the dispersion around the initial $\porb$ values. The peak for the tight population is narrow, since, in our setup, the initial orbital period of [2 3]  is fixed at the median $\porb$ of the observed EW binaries. In contrast, the initial orbital period of [0 1] is taken from a wide distribution (\autoref{sec:method}; \autoref{tab:inprop}), and hence the peak for the wide population is much broader. 

Exchange-ionization can simply be seen as an extension of the double-exchange process. For exchange-ionization that successfully produces an EW--MS binary, we need the regular WD (star 3) to be exchanged from [2 3] with a sufficiently high recoil speed to escape the system completely as a single star, instead of forming a binary with the remaining MS star. Thus, the EW--MS binary cannot absorb sufficient energy to widen. As a result, the exchange-ionization channel contributes only to the tight population of EW--MS binaries.   

The $\porb$ distributions for EW--MS binaries originating from double exchange and exchange-collision are similar, except that exchange-collision outcomes are significantly fewer in number compared to double-exchange outcomes. In the wide population, double-exchange and exchange-collision encounters are also very similar in nature; in the case of exchange-collision, the final tight binary simply has a pericenter distance sufficiently small for a collision. On the other hand, the EW--MS binaries created via exchange-collision in the tight population come from the cases where the wide binary itself has a high eccentricity and small SMA (see, e.g., \autoref{fig:ae}). 
The exchange-collision channel contributes primarily to the wide population. 

\subsection{Rate of formation}
\label{sub:rate}
\begin{figure}
\epsscale{1.2}
    \plotone{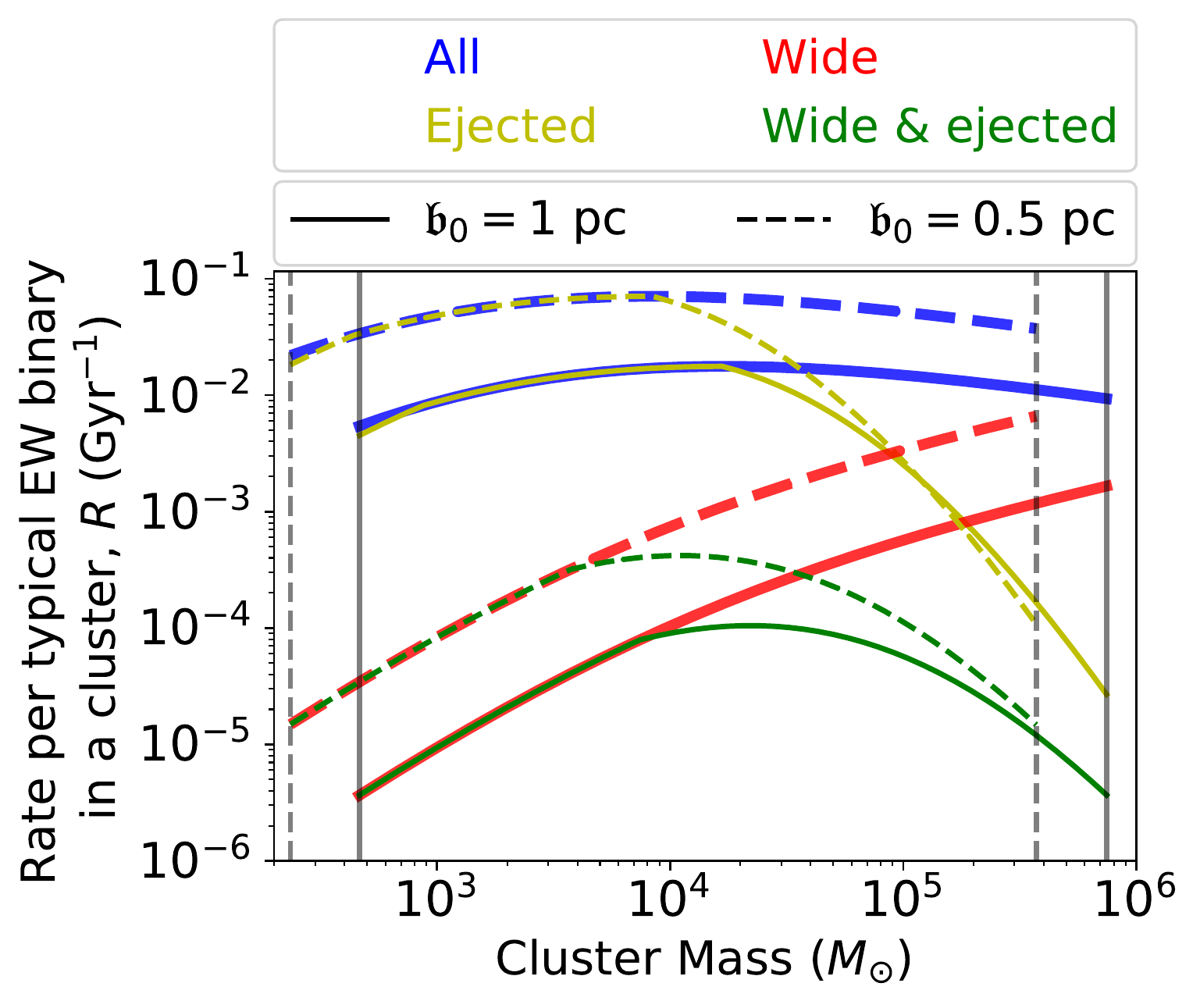}
    \caption{The formation rate, $R$ of EW--MS binaries per typical EW--WD binary in a cluster as a function of cluster mass, $\mcl$. The blue color represents all EW--MS binaries, yellow represents only the ones that are ejected from the host cluster, red represents the ones in wide orbits, and green represents those that are wide as well as ejected. The vertical lines represent the cluster masses corresponding to velocity dispersions, $v_{\sigma}$, of $1$ and $40\,\kms$. This is the range in which we carried out our simulations. Furthermore, we assume star clusters to be Plummer spheres in order to estimate the velocity dispersion, $v_{\sigma}$ as a function of cluster mass, $\mcl$, for plotting this figure. The solid (dashed) colored lines show the rates for the Plummer scale length $\plummerscale/{\rm{pc}} = 1$ ($0.5$), while the vertical lines denote the corresponding simulated range.
    }
    \label{fig:rvsmc}
\end{figure}
The formation rate, $R_i$, of EW--MS binaries from a particular binary-mediated channel $i$ per typical tight EW binary inside a star cluster can be written as:
\begin{equation}
\label{eq:Ri1}
    R_i = n_{\bin} \sigma_i \bar{v},
\end{equation}
where $n_{\bin}$ is the number density of the double-MS binaries inside the cluster, $\sigma_i$ is the cross section of the category $i$, and $\bar{v}$ is the mean relative velocity between the two interacting binaries.
Using $\mcl \sim 2 \rcl v^2/G$ for virialized clusters \citep[$\mcl$, $\rcl$, and $v$ denote the mass, size, and rms speed of the cluster, respectively; e.g.,][]{Wang2020a},
an average stellar mass for the cluster as $\bar{m}$, and replacing $\rcl$ in favor of $\mcl$, we can estimate the stellar number density as
\begin{equation}
\label{eq:numdensity}
    n \sim \frac{\mcl}{\bar{m}} \frac{3}{4 \pi \rcl^3} \sim \frac{6 v^6}{\pi \bar{m} G^3 \mcl^2} .
\end{equation}
The number density for double-MS binaries is then simply 
\begin{equation}
\label{eq:binnumdensity}
    n_{\bin} = f_b n \sim \frac{6 v^6 f_b}{\pi \bar{m} G^3 \mcl^2},
\end{equation}
where $f_b$ is the binary fraction. 

Further, setting $\bar{v}\approx v \approx \sqrt{3}\ v_\sigma$, the rate, $R_i$, can be written as
\begin{equation}
\label{eq:Ri2}
    R_i \sim \frac{162 \sqrt{3} v_{\sigma}^7 \sigma_i f_b}{\pi \bar{m} G^3 \mcl^2}.
\end{equation}
Combining \autoref{eq:Ri2} and \autoref{eq:cseq}, we can estimate the rate of formation per typical tight-orbit EW-WD binary in any cluster with a given $f_b$, $\mcl$, and $v_\sigma$. For example, the rate of formation of wide EW--MS binaries per typical tight EW--WD binary is
\begin{eqnarray}
\Rwide & \sim & 1.56\times10^{-4}  
\left( \frac{\bar{m}}{\msun{}} \right)^{-1} 
\left( \frac{\mcl}{10^5\msun{}} \right)^{-2}
\left( \frac{f_b}{0.5} \right)
\nonumber \\
& \times &
\left( \frac{v_{\sigma}}{10\,\kms} \right)^7  \nonumber \\
& \times &
\left( \frac{v_{\sigma}}{10\,\kms}+0.385 \right)^{-2.25} \rm{Gyr^{-1}}.
\label{eq:Rwide}
\end{eqnarray}

It is instructive to connect the formation rate, $R$ to the host cluster mass, $\mcl$. In a real star cluster, $v_\sigma$ is related to $\mcl$, but for a cluster of a given $\mcl$, since the concentration can vary, so can $v_\sigma$. Nevertheless, in the spirit of this paper, we will make some broad assumptions to estimate an approximate dependence of the rate $R$ on $\mcl$. For simplicity, we treat the clusters as Plummer spheres. In a Plummer sphere, $v_\sigma$ is related to $\mcl$ via the scale length $\plummerscale$: 
\begin{equation}
\label{eq:vplum}    
    v_{\sigma} = \sqrt{\frac{G\mcl}{2\plummerscale}}.
\end{equation}

We also need to adopt $f_b$ which can be dependent on $\mcl$. Observational evidence suggests that $f_b$ is not constant for clusters of all masses. The lower end of the cluster masses, i.e.,  $\mcl/\msun\lesssim10^3-10^4$ corresponds to open clusters with typical values of $f_b\sim0.5$ \citep[e.g.,][]{Jadhav2021}, while on the other extreme, i.e., for $\mcl/\msun\gtrsim10^5$, we have old GCs with a typical $f_b\sim0.05$ \citep[e.g.,][]{Milone2012}. The progenitors of these GCs could possibly have had a higher $f_b$ \citep[e.g.,][]{Leigh2015}. For simplicity, we assume the binary fraction to be $\mcl$-independent and equal to $0.5$.

Finally, we assume $\bar{m}\sim1\,\msun$. In \autoref{fig:rvsmc}, we show the per-target rate of formation $R$ as a function of the cluster mass for all dynamically produced EW--MS binaries (blue), those that are in wide orbits (red), those that are ejected from the host cluster (yellow), and those that are ejected and have wide orbits (green). 
For a fixed $f_b$, while the per-target formation rate for wide EW--MS binaries keeps increasing with cluster mass up to about $\mcl/\msun\sim10^6$, the rate of production of the wide EW--MS binaries that are also ejected from the cluster due to recoil shows a peak near $\mcl/\msun\sim10^4$. This is because the escape speed of star clusters increases with increasing mass, and thus ejection via a single dynamical encounter becomes less likely.

\begin{figure}
\epsscale{1.2}
    \plotone{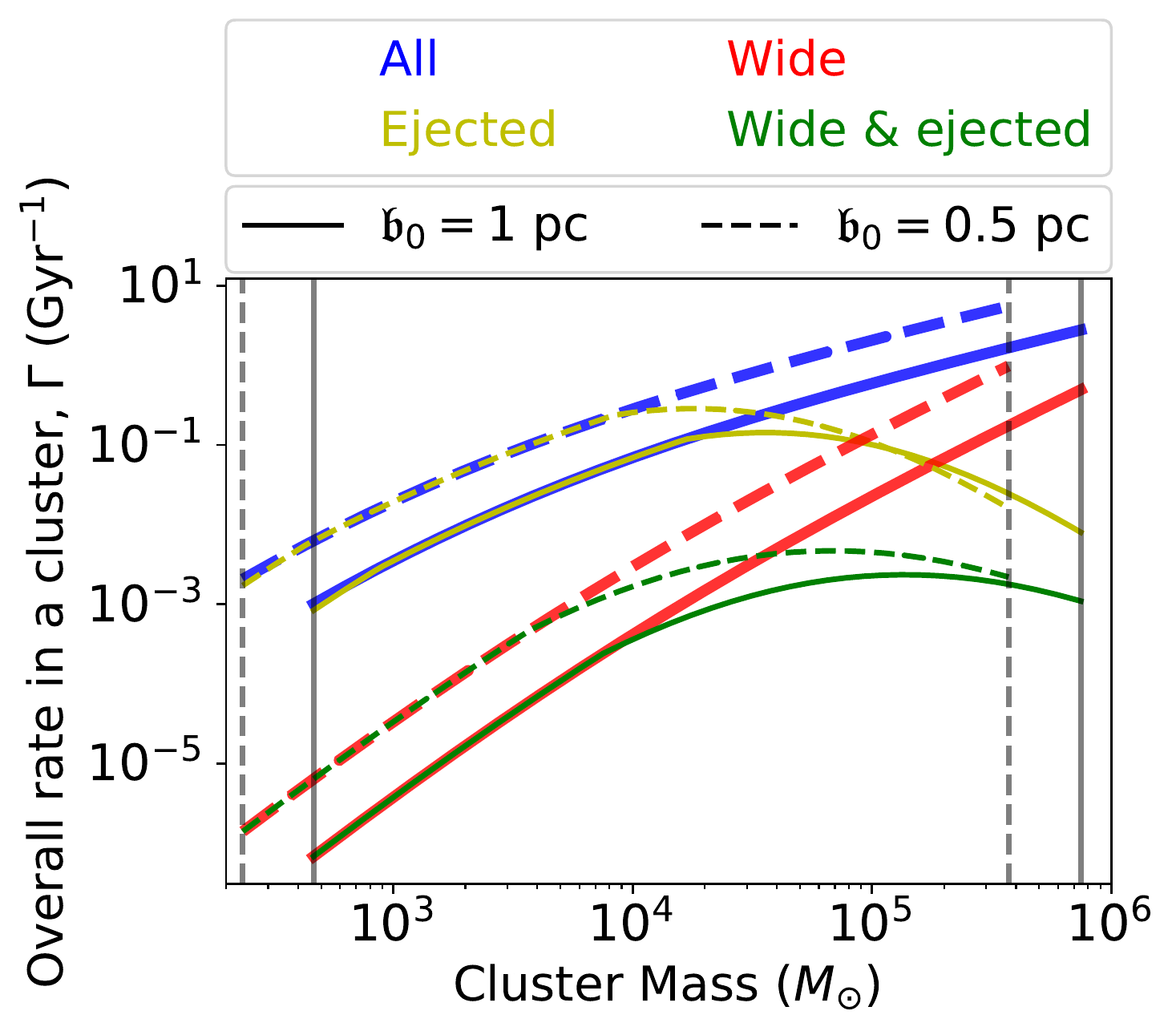}
    \caption{The same as \autoref{fig:rvsmc}, except showing the overall formation rate, $\Gamma$, of EW--MS binaries in a cluster.}
    \label{fig:rvsmc2}
\end{figure}
The overall rate of formation for EW--MS binaries, $\Gamma$ in a star cluster of mass $\mcl$, can be estimated as
\begin{equation}
    \Gamma_i = R_i N_{\rm{EW}},
    \label{eq:gamma}
\end{equation}
where $N_{\rm{EW}}$ is the total expected number of typical tight EW binaries present inside the cluster formed via mass transfer in isolated binary stellar evolution. We estimate $N_{\rm{EW}}$ from binary population synthesis using \cosmic\ \citep{Breivik2020}. We use the same zero-age distributions of binary properties as described in \autoref{sec:popsynth}, and identify double-WD binaries with at least one EW with $\melmwd/\msun\leq0.4$ created within an age of $14\,\gyrs$. We find that this type of binary continues to form throughout the $14\,\gyrs$ of simulation. 
As there is no restriction on the epoch of the binary-binary encounter that created the EW-MS binary, we count the average number of EW-WD binaries existing in the cluster between the ages of $0-14\,\gyrs$. This gives us the average number of EW-WD binaries per simulated binary mass, $n_{\rm{EW}}\approx8\times10^{-4}\,\msun^{-1}$. The total number of EW binaries expected to exist at a particular epoch in a star cluster of mass $\mcl$ and binary fraction $f_b$ is then simply  

\begin{equation}
    N_{\rm{EW}} = n_{\rm{EW}}f_b\mcl.
    \label{eq:New}
\end{equation}
Using \autoref{eq:gamma} and \autoref{eq:New}, we find the total rate of formation of EW--MS binaries:  
\begin{equation}
    \Gamma_i \sim \frac{162 \sqrt{3} v_{\sigma}^7 \sigma_i f_b^2 n_{\rm{EW}}}{\pi \bar{m} G^3 \mcl}.
    \label{eq:Gammai}
\end{equation}

We can now use $\sigma_i(v_\sigma)$, estimated using our scattering experiments (\autoref{eq:cseq}) and \autoref{eq:Gammai} to estimate the total formation rate of EW--MS binaries in a cluster of a given mass.
For example, the formation rate of all EW--MS binaries is
\begin{eqnarray}
    \Gammaall & \sim &
    2.27\times10^{-1}
    \left( \frac{\bar{m}}{\msun{}} \right)^{-1} \left( \frac{\mcl}{10^5\msun{}} \right)^{-1}
    \left( \frac{f_b}{0.5} \right)^{2} \nonumber \\
    & \times &
    \left( \frac{n_{\rm{EW}}}{8\times10^{-4}\,\msun^{-1}} \right)
    \left( \frac{v_{\sigma}}{10\,\kms} \right)^7  \nonumber \\
    & \times &
    \left( \frac{v_{\sigma}}{10\,\kms}+0.133 \right)^{-3.66} \rm{Gyr^{-1}},
\label{eq:Gammaall}
\end{eqnarray}
and the same for wide EW--MS binaries is
\begin{eqnarray}
    \Gammawide & \sim & 6.22\times10^{-3}  
    \left( \frac{\bar{m}}{\msun} \right)^{-1} 
    \left( \frac{\mcl}{10^5\msun} \right)^{-1}
    \left( \frac{f_b}{0.5} \right)^{2} \nonumber\\
    & \times & 
    \left( \frac{n_{\rm{EW}}}{8\times10^{-4}\,\msun^{-1}} \right)
    \left( \frac{v_{\sigma}}{10\,\kms} \right)^7
   \nonumber\\
    & \times & \left( \frac{v_{\sigma}}{10\,\kms}+0.385 \right)^{-2.25} \rm{Gyr^{-1}}.
    \label{eq:Gammawide}
\end{eqnarray}

\autoref{fig:rvsmc2} shows the overall formation rate of EW--MS binaries in a star cluster as a function of $\mcl$. The overall formation rate of EW-MS binaries,  $\Gammaall$, scales strongly with $\mcl$, simply because a more massive cluster would create a larger number of EW binaries (e.g., \autoref{eq:New}), which can be modified to create EW--MS binaries. Similar to $\Rwideejected$, $\Gammawideejected$ shows a peaked distribution, since while a higher $\mcl$ increases the number of EWs the cluster can form, it also increases $\vesc$, making ejections after the binary-binary encounter less likely. We find that $\Gammawideejected$ peaks at $\mcl/\msun\sim 10^5$. 

In order to convert the $\mcl$-dependent formation rate to an overall rate of formation in the Milky Way, we assume an $\mcl$-independent $f_b$, the average fraction of stars that are born in star clusters,\footnote{Some of these clusters dissolve and populate the field \citep[e.g.,][]{Lada2003}.} $\fcl$, a cluster initial mass function given by $d\ncl/d\mcl\propto\mcl^{-2}$ \citep[e.g.,][]{Lada2003}, and the stellar mass estimate of the Milky Way to be $5.4\times10^{10}\,\msun$ \citep{McMillan2017}. It is then straightforward to find an approximate yield for the Milky Way from
\begin{equation}
\label{eq:galacticrate}
    \Gamma^{\rm{MW}}_i = \int_{M_{\rm{cl, min}}}^{M_{\rm{cl, max}}} \Gamma_i \,d\ncl,
\end{equation}
where $M_{\rm{cl, min}}$ ($M_{\rm{cl, max}}$) is the minimum (maximum) cluster mass. The range in $v_\sigma$ used in our scattering experiments corresponds to a range in mass
\begin{eqnarray}
\label{eq:mminmax}
    M_{\rm{cl,min}} &=& \left(\frac{\mathfrak{b_0}}{1\,\rm{pc}}\right) 465\,\msun \nonumber\\
    M_{\rm{cl,max}} &=& \left(\frac{\mathfrak{b_0}}{1\,\rm{pc}}\right) 7.44\times10^5\,\msun
\end{eqnarray}
assuming Plummer spheres (\autoref{eq:vplum}). Using these assumptions, we find that the overall production rate of wide EW--MS binaries in the Milky Way is
\begin{eqnarray}
\label{eq:galacticrate2}
    \Gamma^{\rm{MW}}_{\wide} &\sim& 4.16\times10^3 \left(\frac{\fcl}{0.5}\right)\left(\frac{M_{\rm{MW}}}{5.4\times10^{10}\,\msun}\right) \nonumber \\
    & \times &
    \left(\frac{\mathfrak{b_0}}{1\,\rm{pc}}\right)^{-2} \left( \frac{n_{\rm{EW}}}{8\times10^{-4}\,\msun^{-1}} \right) \nonumber \\
    & \times &
    \left( \frac{f_b}{0.5} \right)^{2}
    \left( \frac{\bar{m}}{1\,\msun} \right)^{-1}
    \rm{Gyr^{-1}}.
\end{eqnarray}
\subsection{Orbital eccentricity}
\label{sec:orbit}
\begin{figure}
    \centering
    \epsscale{1.2}
    \plotone{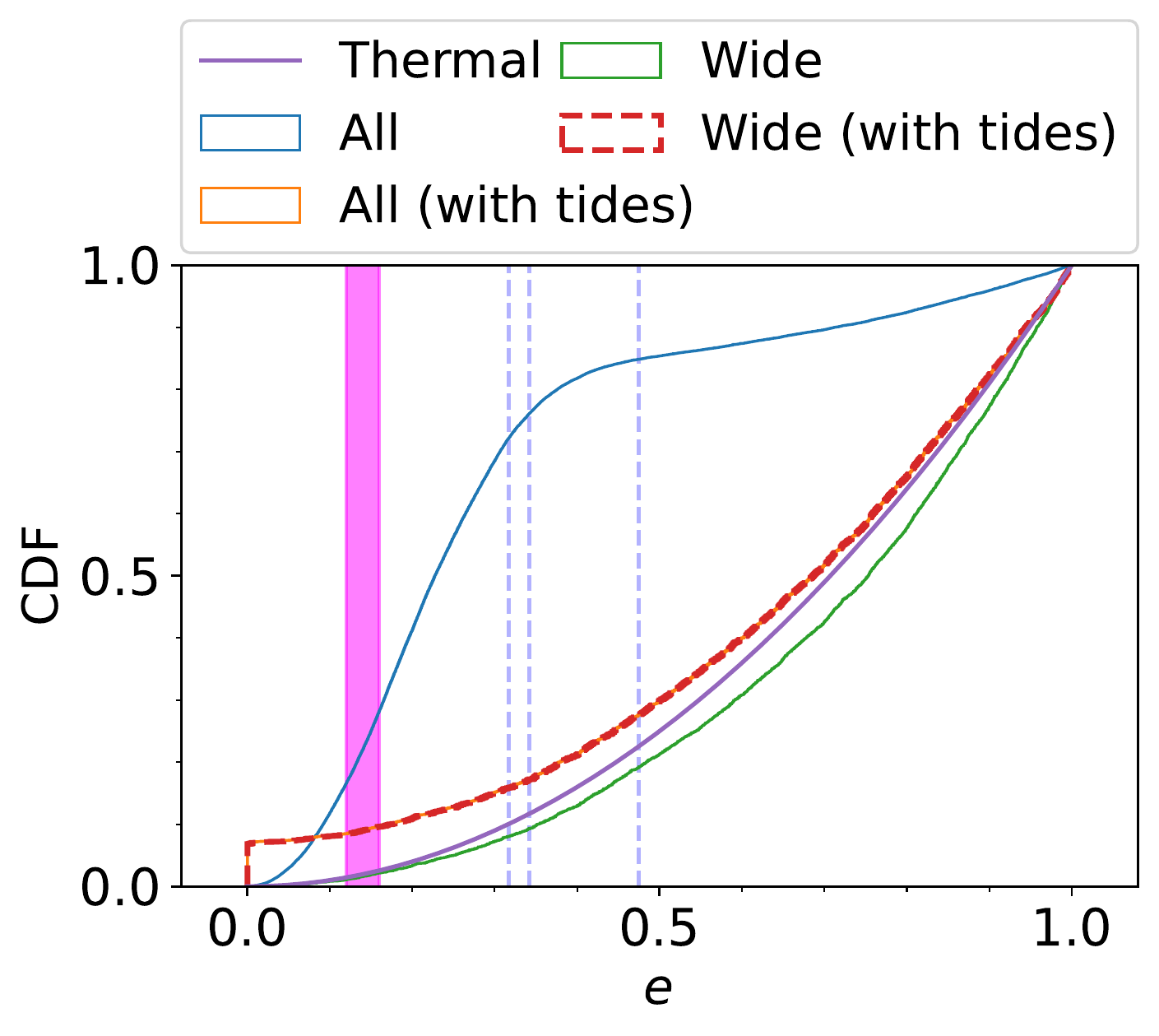}
    \caption{Cumulative distribution of eccentricities for the EW--MS binaries created in our simulations. Blue and green denote all EW--MS binaries and those that are in wide orbits. Orange and the red dashed line denote the expected distribution taking into consideration the tidal evolution for $8\,\gyrs$ post the dynamical production of all and wide binaries, respectively. These two distributions are identical, as almost all the binaries in the tight population merge due to tides. The purple shows the thermal distribution for reference. While it is unknown when the \ab\ may have been created, tidal evolution could not have happened for longer than $\sim8\,\gyrs$ since the MS star in the \ab\ is massive $1.1\,\msun$ \citep{Masuda2019}. The magenta shaded region shows the observed eccentricity within error bars for the \ab{} \citep{Masuda2019}. The blue vertical dashed lines show the eccentricities of the three WD-BSS binaries from \citet{Pandey2021}, which are candidate EW binaries in wide orbits.
    }
    \label{fig:ecchist}
\end{figure}

The observed orbital eccentricity of the \ab\ is low, $\eobs=0.143^{+0.015}_{-0.012}$ (also see \autoref{tab:ab}). On the other hand, dynamically formed systems are expected to have high eccentricities. \autoref{fig:ecchist} shows the eccentricity distribution for the EW--MS binaries created in our simulations. We find that while about $32\%$ of EW--MS binaries can have an eccentricity $e\leq\eobs$, only about $2.5\%$ of the simulated wide EW--MS binaries have $e\leq\eobs$. Using \bse\ \citep{Hurley2002}, we consider the tidal evolution of the EW--MS binaries in wide orbits after dynamical formation, which increases the fraction of wide EW--MS binaries with $e\leq\eobs$ to $10\%$. Thus, indeed, if the \ab\ is created via a binary-binary strong encounter, the low observed eccentricity makes it rare. Based on our simulations, we expect that several more wide EW binaries may exist in higher-eccentricity orbits. Interestingly, \citet{Pandey2021} have indeed reported three EW companions to blue straggler stars in wide orbits (the magenta stars in \autoref{fig:survey}) with relatively higher eccentricities ($e\sim0.475$, $0.317$, and $0.342$) residing in the open cluster M67. We mark these $e$ values by the blue dashed vertical lines in \autoref{fig:ecchist}.

Another implication of our study is the possibility of creating tight EW binaries in eccentric orbits (e.g., see \autoref{fig:ae}). This is relevant for a few interesting candidates, including J1631+0605, an EW binary in a $\approx6\,\hrs$ orbit with a possible modest eccentricity, $e\leq0.3$ \citep{Brown2016}; and the so-called eccentric millisecond pulsars (eMSPs), EW--MSP binaries with eccentricities, $e\sim0.03-0.14$ \citep{2018A&A...612A..78O,2021ApJ...909..161H}. The tight population of simulated EW binaries inherit their orbital energy from the initial short-period [2 3] binary and their orbits become eccentric due to scattering. Note that in general, the EW can have any companion based on the stellar types that constitute the perturbing binary [0 1]. Furthermore, instead of exchange encounters, preservation encounters also excite eccentricities. Preservation encounters are likely more relevant for eMSPs, as it would be highly unlikely for an EW to find an MSP in a random encounter, considering their rarity.

The survival timescales of these eccentric and tight EW binaries under tidal evolution will depend on their orbital properties as well as the companion type. For example, the tight binaries in our simulations having an MS companion all merge within $\sim8$ Gyr (\autoref{fig:ecchist}).

\section{Summary and Discussion}
\label{sec:conclusions}

The observed EW in a wide orbit with a MS star, KIC 8145411 \citep{Masuda2019}, contradicts the expectation that EWs form via mass transfer in an isolated binary \citep[e.g.,][]{Lin2011} and should be in a tight orbit. In this paper, we have analyzed the possibility of dynamically assembling \ab-like wide EW binaries inside star clusters. In particular, we test whether an EW in a tight orbit with another WD, with properties similar to the EW binaries predominantly observed, can be converted to an EW binary similar to the \ab\ via binary-mediated dynamical encounters that are common inside star clusters. We show that, at the least, a binary-binary interaction is required to expand the orbit of the EW binary by the desired amount (\autoref{fig:energy}). Assuming that the typical tight-orbit EW--WD binaries \citep{Brown2016} naturally form via binary stellar evolution, we simulate millions of binary-binary scattering experiments using the small-$N$ body dynamics code, \fewbody\ \citep{Fregeau2004,Fewbodycode2012}, where an EW--WD binary with typical properties is the target binary and another MS--MS binary interacts with it. We carry out these simulations in a variety of star cluster environments, represented by the velocity dispersion, $v_{\sigma}$, ranging from $1$ to $40\ \rm{km\,s^{-1}}$. We collect EW--MS binaries resulting from exchange encounters and study their properties. 

We find that scattering interactions between a typical EW binary and a typical double-MS binary inside the cluster can form EW--MS binaries that are wide enough to resemble the \ab{} and even wider (\autoref{fig:ae}). We find that the dominant channel for creating wide EW--MS binaries is ``nonresonant double exchange" (see \autoref{fig:period1} and \autoref{fig:period2}). The distribution of orbital periods for the EW--MS binaries created in these interactions is bimodal: one mode corresponds to orbits that are tighter than the limiting orbital period for EW binaries expected to form via isolated binary star evolution, and the other is very near the observed orbital period of the \ab. The bimodal distribution of the orbital periods of EW--MS binaries is a direct consequence of the ``double-exchange" process (\autoref{fig:cartoon}), as the final binaries inherit the energies of the initial binaries to a large extent. The natures of the interactions in the other two formation channels, namely ``exchange-collision" and ``exchange-ionization," are very similar to that of double exchange; ``exchange-collision" just requires one of the two final binaries to be close enough at periastron to collide, while ``exchange-ionization" requires the ejected member from the initially tight binary to have a recoil speed large enough such that it does not form a bound state with the remaining MS star. Consequently, ``exchange-collision" also creates a bimodal distribution of orbital periods, while ``exchange-ionization" only creates tight binaries.

Although binary-binary strong encounters inside star clusters do create wide EW--MS binaries, the branching ratio for these outcomes is low ($\lesssim 10^{-3}$ throughout the range of $v_\sigma$ that we have explored; \autoref{fig:csbrl}). Using our simulated scattering experiments, we estimate the cross sections for creating EW--MS binaries via binary-binary interactions as a function of the velocity dispersion of the host star cluster (\autoref{eq:cseq}, \autoref{fig:csbrr}). Simulating $10^6$ binaries using the binary population synthesis code \cosmic\ \citep{Breivik2020}, we estimate the expected number of typical tight EW binaries that form via standard isolated binary evolution per unit mass in a star cluster. We then proceed to estimate the overall rate of production for EW--MS binaries similar to the \ab\ in the Milky Way. Assuming fiducial values for the fraction of stars that form in star clusters, $\fcl=0.5$, $f_b=0.5$, and an initial cluster mass function $\propto \mcl^{-2}$, we estimate that the Milky Way should form wide EW--MS binaries at a rate of $\sim 4.16\times10^3\,\rm{Gyr^{-1}}$ (\autoref{sub:rate}). Furthermore, we find that star clusters in the mass range $\sim 10^4-10^5\,\msun$ are most effective at forming wide EW--MS binaries that also get ejected from the cluster, due to recoil from the binary-binary encounter that created them. 

Note that the Galactic production rate estimated in \autoref{eq:galacticrate2} should be treated as a conservative lower limit because of several restrictions we have imposed. For example, we only count the number of EW-WD binaries in our population synthesis simulation to calculate $N_{\rm{EW}}$, the average number of target EW binaries available for binary-binary scattering (\autoref{eq:New}). In reality, however, other types of EW binaries (e.g., EW-MS, as seen in \autoref{fig:cosmic_kic}, or even EW-neutron stars) could also take part in these interactions and contribute to the formation of EW binaries in wide orbits. Furthermore, to limit the parameter space, our fiducial target binary has been an EW--WD binary with $\porb$ matching the median of the observed EW--WD binaries \citep[$5.4\,\hrs$;][]{Brown2016}. In reality, EW binaries can form in wider orbits, with $\porb$ up to a few days via isolated binary evolution (see, e.g., \autoref{fig:survey} and \autoref{fig:cosmic_kic}). Considering targets in initially wider orbits would proportionally increase the cross sections and likelihood of exchange. On the other hand, if a broader distribution of initial tight EW--WD binaries (instead of simply using the median) were used, the narrow peak of the tight binaries in the orbital period distribution (\autoref{fig:period1} and \autoref{fig:period2}) would also widen. 

While, binary-binary encounters inside star clusters can naturally produce EW--MS binaries in orbits wider than the \citet{Lin2011} boundary, starting from typical EW--WD binaries, the EW--MS binaries produced through this channel typically have relatively high eccentricities (\autoref{fig:ae}). Due to the wide orbits, these eccentricities may be damped only in a small fraction of these binaries after formation. We find that up to about $10\%$ of all wide EW--MS binaries produced via binary-binary encounter may have eccentricities similar to the \ab\ or lower. Thus, if indeed EW binaries in wide orbits do form via dynamical encounters, it is expected that more wide-orbit EW binaries may be there in eccentric orbits \citep[e.g.,][]{Pandey2021}.

Since the \ab\ is observed in the field, throughout the manuscript we have studied production rates for wide EW binaries, as well as a subset of those with recoils expected to eject them from the host cluster. The latter subset should be thought of as the lower limit of production, since, in principle, the widening may happen via multiple encounters instead of a single encounter. Similarly, even if not ejected from the host cluster, the dynamically created wide EW binary may simply be lost from the cluster across the cluster's tidal radius. Moreover, if created in a low-mass cluster, the host cluster itself may dissolve completely at an epoch.   

In this study we have explored a plausible way of forming EW binaries with orbits that are wider than what is theoretically allowed and typically observed, without any need to modify our present understanding of stellar binary interactions. We have found that binary-binary interactions between typical short-period EW--WD and MS--MS binaries can create wide EW binaries with similar properties to the \ab. Furthermore, we find that systems similar to the \ab\ must be rare.

\begin{acknowledgments}

We thank the anonymous referee for very helpful suggestions and comments. A.K. and C.C. acknowledge support from TIFR's graduate fellowship. S.C. acknowledges support from the Department of Atomic Energy, Government of India, under project no.  12-R\&D-TFR-5.02-0200 and RTI 4002. All simulations were done using the TIFR HPC.

\end{acknowledgments}

\software{\fewbody\ \citep{Fregeau2004,Fregeau2006}; \cosmic\ \citep{Breivik2020}; \bse\ \citep{Hurley2002}; \texttt{matplotlib}\ \citep{matplotlib}; \texttt{numpy}\ \citep{numpy}; \texttt{scipy}\ \citep{scipy}; \texttt{pandas}\ \citep{pandas}; \texttt{seaborn}\ \citep{Waskom2021}}

\bibliographystyle{aasjournal}

\end{document}